# Phonons, electronic charge response and electron-phonon interaction in the high-temperature superconductors

Claus Falter

Institut für Festkörpertheorie, Universität Münster,

Wilhelm-Klemm-Str. 10, 48149 Münster, Germany

E-mail: Falter@nwz.uni-muenster.de

**Abstract**

We investigate the complete phonon dispersion, the phonon induced electronic charge response and the corresponding self-consistent change of the crystal potential an electron feels as a direct measure of the electron-phonon interaction in the high-temperature superconductors within a microscopic model in the framework of linear response theory. Moreover, dielectric and infrared properties are calculated. The experimentally observed strong renormalization of the in-plane oxygen bond-stretching modes which appears upon doping in the high-temperature superconductors is discussed. It is shown that the characteristic softening, indicating a strong nonlocal electron-phonon interaction, is most likely a generic effect of the *CuO* plane and is driven by a nonlocal coupling of the displaced ions to the localized charge-fluctuations at the *Cu* and the $O_{xy}$ ions. At hand of the oxygen bond-stretching modes it is illustrated how lattice-, charge- and spin-degrees of freedom may act synergetically for anisotropic pairing in the high-temperature superconductors. The different behaviour of these modes during the insulator-metal transition via the underdoped phase is calculated and from a comparison of these generic modes in the different phases conclusions about the electronic state are drawn. For the non-cuprate potassium doped high-temperature superconductor *Ba-Bi-O* also a very strong and anisotropic renormalization of the oxygen bond-stretching modes is predicted. In another investigation *c*-axis polarized infrared- and Raman-active modes of the HTSC's are calculated in terms of charge fluctuations and anisotropic dipole-fluctuations. Mode assignments discussed controversially in the literature are proposed. Finally, interlayer phonons propagating along the *c*-axis and their accompanying charge response are investigated. Depending on the strength of the interlayer coupling calculations are performed ranging from the static, adiabatic response regime to the non-adiabatic regime where dynamical screening of the bare Coulomb interaction and phonon-plasmon coupling becomes relevant within a certain region around the direction of the *c*-axis. A comparison with the experimental situation is given. Both, the oxygen bond-stretching modes calculated in adiabatic approximation and the non-adiabatic coupled *c*-axis phonon-plasmon modes are found to be important for pairing via lattice- and charge-degrees of freedom. Favouring aspects to achieve high-temperature superconductivity are also discussed.

## 1. Introduction

There are numerous experimental observations that electron-phonon interaction (EPI) plays a nontrivial role in the optical, transport, lattice dynamical and superconducting properties of the cuprate based high-temperature superconductors (HTSC's) [1, 2]. In particular, inelastic neutron scattering experiments [3 - 6] have shown that the in-plane high-frequency oxygen bond-stretching modes (OBSM) are strongly renormalized upon doping indicating a strong EPI in the HTSC's. Evidence that these modes should play a role for superconductivity also comes from tunnelling spectroscopy [7] and penetration depth measurements [8]. Recently, substantial interest in the phonons and the EPI resulted from angle-resolved photoemission spectroscopy which suggests a strong and ubiquitous electron-phonon coupling [9, 10]. Large and anomalous oxygen isotope effects in particular in the underdoped regime also provide evidence for the important role of the phonons, see e. g. Refs. [11, 12]. Moreover, theories predicting charge inhomogeneities in the *CuO* plane in form of stripes [13 – 16] have attracted interest in the phonons because such charge distributions will inevitably couple to the lattice. So, certain lattice distorsions, like the OBSM, could play a role in supporting dynamic charge inhomogeneities through the induced charge-fluctuations. A general review on the role which phonons play in the HTSC's can be found in [17]. Altogether, from the experimental evidence we conclude that EPI is an important aspect for the physics in the HTSC's.

Most ab initio calculations of the phonon frequencies and the EPI in the HTSC's have been performed for single modes of high symmetry type only, using the frozen phonon technique within density functional theory (DFT) and applying the local density approximation (LDA), see e. g. Refs. [18 – 25]. Only very few attempts have been made to calculate the complete phonon dispersion and the EPI using the linear response method in form of density-functional perturbation theory (DFPT) within LDA [26 – 28]. From these first-principles calculations no unique conclusion can be drawn concerning the magnitude of the EPI in the HTSC's. The results scatter between large coupling strengths for specific modes and a relatively small overall coupling averaged over the Brillouin zone (BZ).

However, when calculating the EPI along the (static) DFT-LDA ground-state-route one should have in mind that e.g. possible effects from strong electron correlation and in particular from a non-adiabatic *dynamical* charge response are missing. When compared to LDA correlation effects produce besides damping a smaller bandwidth and smaller electron velocities, see e. g. [29]. This leads to corrections of the single-particle contribution to the density response function. Moreover, and consistent with the latter corrections, the exchange-correlation part is changed by strong electron-electron interaction. A non-adiabatic charge response in a certain region of reciprocal space around the $c$-axis of the HTSC's becomes very likely because of the large anisotropy of these materials leading to a very weak interlayer coupling and consequently a strongly reduced screening [30 – 32]. As a consequence, the time scale of the electron- and phonon-dynamics around the $c$-axis is of the same order. Thus, the adiabatic approximation breaks down in the metallic phase and the dynamical nature of the poorly screened Coulomb interaction becomes important. DFT-LDA calculations, which are based on the adiabatic approximation consistent with static screening, underestimate the anisotropy considerably [18, 33] and simulate a charge response of a less anisotropic three dimensional metal. So, calculating the phonon dispersion and the EPI around the $c$-axis within (static) DFT becomes questionable. Finally, DFT-LDA calculations always find a metallic groundstate and consequently can deal only with the doped, metallic phase of the HTSC's outside the non-adiabatic region. Thus, a theoretical comparison of the vibronic properties with those of the corresponding insulating parent compounds or the underdoped phase of the HTSC's is not possible.

In our microscopic model approach to the HTSC's the charge response in the metallic and insulating phase within the adiabatic approximation as well as the non-adiabatic charge response around the $c$-axis in the metal can be calculated, at least approximatively. A survey of the results obtained with our theoretical method is presented in Section 3. In this approach all the couplings appearing in the dynamical matrix and the EPI are microscopically well defined and can be calculated. Of course, the method needs some analytical input and makes some physical motivated but controlled approximations which, however, are well suited for the HTSC's as materials with a dominating component of ionic binding and localized charge distributions. Anyway, the numerical calculations are far less time consuming than for a canonical procedure in the DFT-framework. Ultimately, this allows also for investigations in the non-adiabatic response regime. By varying certain microscopic well defined coupling coefficients of the theory as compared to their calculated values, like the important on-site Coulomb interaction at the copper or the dipole polarizability of the ions, their specific effect on the properties to be calculated, like the phonon anomalies, can be studied directly in numerical simulations.

In our microscopic modelling of the charge response and lattice dynamics in the framework of the linear response approach we focus on the specific electronic features of the cuprate superconductors. The ionic nature of the HTSC's is described by an ab initio rigid-ion model (RIM) leading to a local, rigid charge response and EPI, respectively, while the nonlocal, non-rigid part of the electronic density response and EPI is modelled by microscopically well defined charge- and dipole-fluctuations (CF's and DF's) localized on the outer shells of the ions because most of the charge in the HTSC's contributing to bonding is concentrated on the latter. Additionally, covalent metallic features of bonding in the HTSC's are approximatively taken into account by the electronic bandstructure : first in a global way by using (static) effective ionic charges in the RIM (ion softening), which can be calculated from a tight-binding analysis of a given bandstructure, and second, via the electronic polarizability, which contains the kinetic single-particle part of the electronic density response. The latter can be generalized from its static form consistent with the adiabatic approximation to its dynamical version which is relevant for a description of phonon-plasmon mixing in the non-adiabatic regime. An appropriate model can be constructed for the insulating phase of the HTSC's in a more semiempirical way by taking a sum rule strictly valid for the electronic density response of an insulator into account being qualitatively different from that of a metal because of the energy gap in the excitation spectrum of the charges. Finally, for the underdoped phase of hole-doped materials a model is proposed suitable to describe a novel metallic phase with an insulator-like charge response at the $Cu$ sublattice separated from a metallic charge response on the oxygen network of the $CuO$ plane reflecting partial ordering of the conducting carriers in a subspace of the two-dimensional space. Moreover, a modelling of the electron-doped materials expressing electron-hole asymmetry is proposed.

In section 2 a résumé of the theory and modelling is given to provide a better reading of the review. Section 3 is divided up into four parts and presents a survey of our calculated results and more general aspects of the physics in the HTSC's. The first Subsection shows calculations of the phonon dispersion of $La_2CuO_4$ in the metallic and insulating phase and provides a comparison with the experiments. Additionally, dielectric and infrared properties are studied. Next infrared and Raman-active modes of $Bi$-based cuprate superconductors are discussed. In the third Subsection the presumedly generic anomalous phonon softening of the OBSM upon doping is investigated. Moreover, the electronic charge response and the corresponding self-consistent change of the potential an electron feels in these modes is calculated and our modelling of the electronic state of the HTSC's is discussed in some detail. The final Subsection reviews our investigations in the non-adiabatic response regime and explores phonon-plasmon mixing and non-adiabatic coupling effects of the electrons and lattice vibrations. Section 4 contains a summary and the conclusions.

## 2. Résumé of the theory and modelling

In the following a survey of the theory and modelling of our microscopic approach of the electronic density response, the lattice dynamics and the EPI in the HTSC's is given. A detailed description can be found in [34, 35]. In this formulation the local part of the electronic charge response is approximated by an ab initio RIM taking into account covalent ion softening in terms of (static) effective ionic charges which can be calculated from a tight-binding analysis (TBA) of a given single-particle bandstructure. Moreover, scaling of the short-range part of certain pair potentials between the ions is considered in order to simulate additional covalence effects in the calculations [36]. Scaling is performed in such a way that the energy-minimized structure is as close as possible to the experimental one. Note, that structure optimization and energy minimization is very important for a reliable calculation of the phonon dynamics through the dynamical matrix. The TBA supplies the effective ionic charges as extracted from the orbital occupation numbers $Q_\mu$ of the $\mu$ (tight-binding) orbital in question:

$$Q_\mu = \frac{2}{N} \sum_{nk} \left| C_{\mu n}\left(\vec{k}\right) \right|^2 \quad . \qquad (1)$$

$C_{\mu n}\left(\vec{k}\right)$ denotes the $\mu$ component of the eigenvector of band $n$ at wavevector $\vec{k}$ in the first BZ; the summation in Eq. (1) runs over all occupied states and $N$ gives the number of elementary cells in the (periodic) crystal. The RIM with the corrections just mentioned then serves as an unbiased reference system for the description of the HTSC's and can be considered as a first approximation for the insulating state of the HTSC's leaving spin degrees of freedom aside. Starting from such an unprejudiced rigid reference system non-rigid electronic polarization processes are introduced in form of more or less localized electronic CF's at the outer shells of the ions. Especially in the metallic phase of the HTSC's the latter dominate the nonlocal contribution of the electronic density response and the EPI and are particularly important in the CuO planes. In addition, anisotropic DF's are admitted within our approach [35, 37] which prove to be specificly of interest for the ions in the ionic layers mediating the dielectric coupling. Thus, the starting point of our model is the ionic density in the perturbed state which is given by

$$\rho_\alpha\left(\vec{r}\right) = \rho_\alpha^0\left(r\right) + \sum_\lambda Q_\lambda \rho_\lambda^{CF}\left(r\right) + \vec{p}_\alpha \cdot \hat{r} \, \rho_\alpha^D\left(r\right) \qquad . \qquad (2)$$

$\rho_\alpha^0$ is the density of the unperturbed ion as used in the RIM localized at the sublattice $\alpha$ of the crystal and following the latter rigidly under displacements. The $Q_\lambda$ and $\rho_\lambda^{CF}$ describe the amplitudes and the form-factors of the CF's and the last term in Eq. (2) represents the dipolar deformation of an ion $\alpha$ with amplitude (dipole moment) $\vec{p}_\alpha$ and a radial density distribution $\rho_\alpha^D$. $\hat{r}$ is the unit vector in the direction of $\vec{r}$. The $\rho_\alpha^{CF}$ are approximated by a spherical average of the orbital densities of the ionic shells calculated in LDA taking self-interaction effects (SIC) into account. The dipole density $\rho_\alpha^D$ is obtained from a modified Sternheimer method in the framework of LDA-SIC [35].

The total energy of the crystal is investigated by assuming that the density can be approximated by a superposition of overlapping densities of the individual ions, $\rho_\alpha$. The $\rho_\alpha^0$ are calculated within LDA-SIC taking environment effects, via a Watson sphere potential, and the calculated static effective charges of the ions into account. Such an approximation holds well in the HTSC's [36, 38]. The static effective charges represent a global covalence effect in the system. As a general rule, partial covalence reduces the amplitude of the static effective charges in mixed ionic-covalent compounds like the HTSC's, because the charge transfer from the cations to the anions is not complete as in the entirely ionic case. Moreover, applying the pair-potential approximation we get for the total energy:

$$E\left(R,\zeta\right)=\sum_{\bar{a}\alpha}E_\alpha^{\bar{a}}\left(\zeta\right)+\frac{1}{2}\sum_{\substack{\bar{a}\alpha \\ \bar{b}\beta}}{}'\Phi_{\alpha\beta}\left(\vec{R}_\beta^{\bar{b}}-\vec{R}_\alpha^{\bar{a}},\zeta\right)\qquad.\qquad(3)$$

The energy E depends on the configuration of the ions $\{R\}$ and the electronic (charge) degrees of freedom (EDF) $\{\zeta\}$ of the charge density, i. e. $\{Q_\lambda\}$ and $\{\vec{p}_\alpha\}$ in Eq. (2). $E_\alpha^{\bar{a}}$ are the energies of the single ions. $\vec{a},\vec{b}$ denote the elementary cells in the crystal and $\alpha,\beta$ the corresponding sublattices. The second term in Eq. (3) is the interaction energy of the system, expressed in terms of the *anisotropic* pair-interactions $\Phi_{\alpha\beta}$. The prime in (3) means that the self-term has to be omitted. Both $E_\alpha^{\bar{a}}$ and $\Phi_{\alpha\beta}$ in general depend upon $\zeta$ via $\rho_\alpha$.

From the adiabatic condition

$$\frac{\partial E\left(R,\zeta\right)}{\partial \zeta}=0\qquad,\qquad(4)$$

an expression for the atomic force constants, and accordingly, the dynamical matrix in the harmonic approximation can be derived.

$$t_{ij}^{\alpha\beta}\left(\vec{q}\right)=\left[t_{ij}^{\alpha\beta}\left(\vec{q}\right)\right]_{RIM}-\frac{1}{\sqrt{M_\alpha M_\beta}}\sum_{\kappa,\kappa'}\left[B_i^{\kappa\alpha}\left(\vec{q}\right)\right]^*\left[C^{-1}\left(\vec{q}\right)\right]_{\kappa\kappa'}B_j^{\kappa'\beta}\left(\vec{q}\right).\qquad(5)$$

$\left[t_{ij}^{\alpha\beta}\left(\vec{q}\right)\right]_{RIM}$ denotes the contribution of the RIM to the dynamical matrix. $M_\alpha,M_\beta$ are the masses of the ions and $\vec{q}$ is a wave vector from the first BZ.

The quantities $\vec{B}(\vec{q})$ and $C(\vec{q})$ in Eq. (5) represent the Fourier transforms of the coupling coefficients as calculated from the energy,

$$B_{\kappa\beta}^{\vec{a}\vec{b}} = \frac{\partial^2 E(R,\zeta)}{\partial \zeta_\kappa^{\vec{a}} \partial \vec{R}_\beta^{\vec{b}}} \qquad , \qquad (6)$$

$$C_{\kappa\kappa'}^{\vec{a}\vec{b}} = \frac{\partial^2 E(R,\zeta)}{\partial \zeta_\kappa^{\vec{a}} \partial \zeta_{\kappa'}^{\vec{b}}} \qquad . \qquad (7)$$

The derivatives in Eqs. (6) and (7) have to be performed at the equilibrium positions. $\kappa$ denotes the EDF (CF's and DF's in the present theory) in an elementary cell of the crystal. The $\vec{B}$ coefficients describe the coupling between the EDF and the displaced ions (bare electron-phonon coupling), and the $C$ coefficients the interaction between the EDF.

The phonon frequencies $\omega_\sigma(\vec{q})$ and the eigenvectors $\vec{e}^\alpha(\vec{q}\sigma)$ of the modes $(\vec{q}\sigma)$ then are obtained from the secular equation for the dynamical matrix given in Eq. (5), i. e.

$$\sum_{\beta j} t_{ij}^{\alpha\beta}(\vec{q}) e_j^\beta(\vec{q}) = \omega^2(\vec{q}) e_i^\alpha(\vec{q}) \qquad . \qquad (8)$$

Equations (5) – (8) are generally valid and, in particular, are independent of our specific model for the decomposition of the perturbed density in Eq. (2) and of the pair approximation (3) for the energy.

The interaction $\Phi_{\alpha\beta}$ can be decomposed into long-ranged Coulomb contributions and short-ranged terms. The former can be dealt with by Ewald-techniques and the latter are separated into the interaction between the ion cores and the charge density from Eq. (2), the interaction between the density $\rho_\alpha$ with the density $\rho_\beta$ (Hartree contribution) and a term representing the sum of the kinetic single-particle and the exchange-correlation contribution of the interaction between the ions. A detailed description of the $\Phi_{\alpha\beta}$ and the calculation of the coupling coefficients $\vec{B}$ and $C$ for the EDF is given in [35]. In this context we remark that the matrix $C_{\kappa\kappa'}(\vec{q})$ of the EDF-EDF interaction whose inverse appears in Eq. (5) for the dynamical matrix can be written as

$$C = \Pi^{-1} + \tilde{V} \qquad . \qquad (9)$$

$\Pi^{-1}$ contains the kinetic single-particle part to the interaction and $\tilde{V}$ the Hartree and exchange-correlation contribution. The quantity $C^{-1}$ needed for the calculations of the dynamical matrix is closely related to the (linear) density response function (matrix) and to the inverse dielectric function (matrix) $\varepsilon^{-1}$, respectively. Coming back to the first principles investigations within DFPT [26 – 28] these calculations correspond to calculating $\Pi$ and $\tilde{V}$ in DFT-LDA. On the other hand, in our microscopic modelling DFT-LDA-SIC calculations are performed for the various densities in Eq. (2) in order to obtain the coupling coefficients $\vec{B}$ and $\tilde{V}$. Written in matrix notation we have the relation

$$C^{-1} = \Pi \left( 1 + \tilde{V}\Pi \right)^{-1} \equiv \Pi \varepsilon^{-1} \quad , \quad \varepsilon \equiv 1 + \tilde{V}\Pi \qquad . \qquad (10)$$

The CF-CF submatrix of the matrix $\Pi$ can be calculated approximatively from a TBA of a single-particle electronic bandstructure. In this case the electronic polarizability $\Pi$ is given by

$$\Pi_{\kappa\kappa'}(\vec{q}) = -\frac{2}{N} \sum_{\substack{nn' \\ \vec{k}}} \frac{f_{n'}(\vec{k}+\vec{q}) - f_n(\vec{k})}{E_{n'}(\vec{k}+\vec{q}) - E_n(\vec{k})} \Big[ C^*_{\kappa n}(\vec{k}) C_{\kappa n'}(\vec{k}+\vec{q}) \Big] \Big[ C^*_{\kappa' n}(\vec{k}) C_{\kappa' n'}(\vec{k}+\vec{q}) \Big]^* . \quad (11)$$

$f$, $E$ and $C$ in Eq. (11) are the occupation numbers, the single-particle energies and the expansion coefficients of the Bloch-functions in terms of the tight-binding functions. The self-consistent change of an EDF on an ion generated by a phonon mode $(\vec{q}\sigma)$ with frequency $\omega_\sigma(\vec{q})$ and eigenvector $\vec{e}^\alpha(\vec{q}\sigma)$ can be expressed in the form

$$\delta\zeta_\kappa^{\vec{a}}(\vec{q}\sigma) = \left[ -\sum_\alpha \vec{X}^{\kappa\alpha}(\vec{q}) \vec{u}_\alpha(\vec{q}\sigma) \right] e^{i\vec{q}\cdot\vec{R}_\kappa^{\vec{a}}} \equiv \delta\zeta_\kappa(\vec{q}\sigma) e^{i\vec{q}\cdot\vec{R}^{\vec{a}}} \quad , \qquad (12)$$

with the ionic displacements

$$\vec{u}_\alpha^{\vec{a}}(\vec{q}\sigma) = \left( \frac{\hbar}{2M_\alpha \omega_\sigma(\vec{q})} \right)^{1/2} \vec{e}^\alpha(\vec{q}\sigma) e^{i\vec{q}\cdot\vec{R}^{\vec{a}}} \equiv \vec{u}_\alpha(\vec{q}\sigma) e^{i\vec{q}\cdot\vec{R}^{\vec{a}}} \quad , \qquad (13)$$

where the self-consistent response per unit displacement of the EDF is obtained in linear response theory from

$$\vec{X}(\vec{q}) \equiv \Pi(\vec{q}) \varepsilon^{-1}(\vec{q}) \vec{B}(\vec{q}) = C^{-1}(\vec{q}) \vec{B}(\vec{q}) \qquad . \qquad (14)$$

A measure of the strength of the EPI in a certain phonon mode $(\vec{q}\sigma)$ is provided by the change of the self-consistent potential in the crystal felt by an electron at space point $\vec{r}$ in this mode, $\delta V_{eff}(\vec{r}, \vec{q}\sigma)$. Weighting this quantity with the corresponding density form factor $\rho_\kappa(\vec{r} - \vec{R}_\kappa^{\vec{a}})$ of the EDF located at $\vec{R}_\kappa^{\vec{a}}$, we obtain

$$\delta V_\kappa^{\vec{a}}(\vec{q}\sigma) = \int dV \rho_\kappa(\vec{r} - \vec{R}_\kappa^{\vec{a}}) \delta V_{eff}(\vec{r}, \vec{q}\sigma) \qquad (15)$$

as a direct measure for the strength of the EPI in the mode $(\vec{q}\sigma)$ mediated by the EDF considered. Ultimately $\delta V_\kappa^{\vec{a}}(\vec{q}\sigma)$ can be expressed by the coupling coefficients from Eq. (6) and (7) in the following way:

$$\delta V_\kappa^{\vec{a}} (\vec{q}\sigma) = \left[ \sum_\alpha \vec{W}^{\kappa\alpha} (\vec{q}) \cdot \vec{u}_\alpha (\vec{q}\sigma) \right] e^{i\vec{q}\cdot\vec{R}_\kappa^{\vec{a}}} \equiv \delta V_\kappa (\vec{q}\sigma) e^{i\vec{q}\cdot\vec{R}^{\vec{a}}} \qquad , \qquad (16)$$

where $\vec{W}(\vec{q})$ is the Fourier transform of

$$\vec{W}_{\kappa\beta}^{\vec{a}\vec{b}} \equiv \vec{B}_{\kappa\beta}^{\vec{a}\vec{b}} - {}^{kin}\vec{B}_{\kappa\beta}^{\vec{a}\vec{b}} - \sum_{\vec{b}'\kappa'} \tilde{V}_{\kappa\kappa'}^{\vec{a}\vec{b}'} \vec{X}_{\kappa'\beta}^{\vec{b}'\vec{b}} \qquad . \qquad (17)$$

$^{kin}\vec{B}$ in Eq. (17) denotes the contribution of the kinetic energy to the coupling coefficient $\vec{B}$ in Eq. (6).

The generalization for the quantity $\Pi$ in Eqs. (9) or (10) needed in the non-adiabatic regime, where dynamical screening effects must be considered, can be achieved by adding $(\hbar\omega + i\eta)$ to the differences of the single-particle energies in the denominator of the expression for $\Pi$ in Eq. (11). Other non-adiabatic contributions besides that of the single-particle part to the electronic density response generated by dynamical exchange-correlation effects and the phonons themselves are beyond the scope of the present article. Using Eq. (10) for the density response function or the dielectric function, respectively, and the frequency-dependent version of the polarizability $\Pi$ according to Eq. (11), the free-plasmon dispersion is obtained from the relation

$$\det\left[ \varepsilon_{\kappa\kappa'} (\vec{q}, \omega) \right] = 0 \qquad . \qquad (18)$$

The coupled-mode frequencies of the phonons and the plasmon must be determined self-consistently from the secular equation for the dynamical matrix in Eq. (5) which now contains the frequency $\omega$ implicitly via $\Pi$ in the response function $C^{-1}$. Analogously, the dependence on the frequency is transferred to the quantity $\vec{X}$ in Eq. (14) and thus to $\delta\zeta_\kappa$ and $\delta V_\kappa$ in Eq. (12) and (16), respectively.

## 3. Results and discussion

### 3.1. Phonon Dispersion of $La_2CuO_4$ in the metallic and insulating phase – Dielectric properties.

For an unbiased discussion of the phonon renormalization induced by non-rigid, nonlocal EPI effects of the ionic CF and DF-type mediated by the second term of the dynamical matrix in Eq. (5), a quantitative and unprejudiced reference model, which does not include already the electronic polarization effects to be investigated, is needed. Such a model is provided by the ionic ab-initio RIM introduced in Section 2, representing approximately the local EPI effects.

In this model ion softening, resulting in static effective ionic charges, is calculated from a TBA of the first-principles electronic bandstructure of $La_2CuO_4$ as given in [39] leading altogether to an 31-band-model (31BM). The orbital occupation numbers in Eq. (1) are calculated from this bandstructure and so we obtain the following electron configuration [36]:

$Cu\ 3d^{9.24}4s^{0.3}4p^{0.24}$; $O_{xy}2p^{5.42}$; $O_z2p^{5.47}$ and $La\ 5d^{0.72}$. Correspondingly, the following (static) effective ionic charges can be extracted, $Cu^{1.22+}$, $O_{xy}^{1.42-}$, $O_z^{1.47-}$, $La^{2.28+}$, reflecting partial covalence effects in the HTSC's. A possible (weak) dependence of these charges on doping has not been considered. This could be achieved for example by relying upon experimental results for certain electronic and lattice dynamical properties.

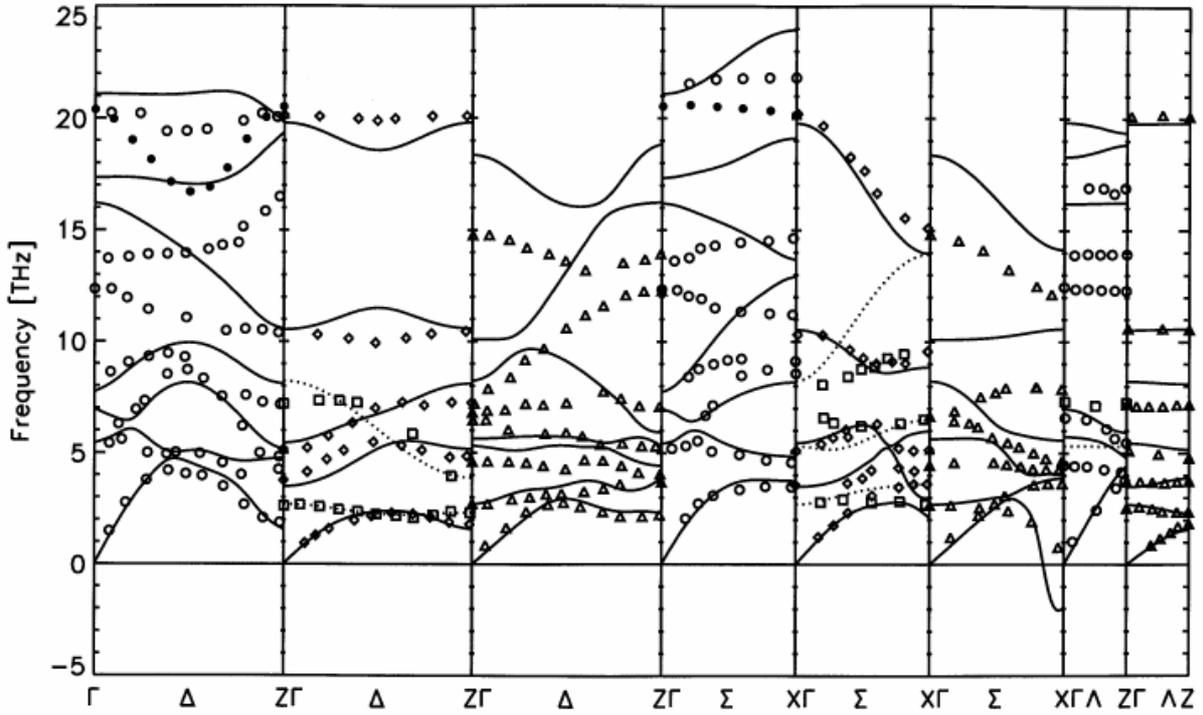

**Fig. 1** Calculated phonon dispersion of tetragonal $La_2CuO_4$ in the main symmetry directions

$\Delta \sim (1,0,0), \Sigma \sim (1,1,0)$ and $\Lambda \sim (0,0,1)$ as calculated from the RIM. The experimental data are taken from Refs. [3, 4, 41]. The various open symbols representing the experimental results for the insulating phase indicate different irreducible representations (ID). The arrangement of the panels from left to right is as follows:

$\Delta_1; (\Delta_2 \cdots; \Delta_4 -), \Delta_3; \Sigma_1; (\Sigma_2 \cdots; \Sigma_4 -); \Sigma_3; (\Lambda_1 -; \Lambda_2 \cdots); \Lambda_3$. The solid circles represent the experimental values of the highest $\Delta_1$ and $\Sigma_1$ branch for the optimally doped, metallic phase $La_{1.85}Sr_{0.15}CuO_4$. As compared with the notation used in the experiments for the classifying ID's the following changes should be noted:

$\Delta_3 \leftrightarrow \Delta_4; \Sigma_3 \leftrightarrow \Sigma_4; \Lambda_3 \leftrightarrow \Lambda_5$. The frequencies are in *THz* and the imaginary frequencies are represented as negative numbers.

The structural parameters in the RIM as calculated from an energy minimization taking scaling into account are in very good agreement with the experimental values:

$$a = 3.784 \overset{o}{A}(3.790) \quad , \quad c = 13.224 \overset{o}{A}(13.227),$$

$$Z(O_z) = 0.1846(0.1820) \quad , \quad Z(La) = 0.1337(0.1380).$$

The experimental data [40] are given in the brackets. The coordinates $Z(O_z)$ and $Z(La)$ are in units of $c$; $a$, $c$ are the planar and axial lattice constants, respectively. Such a good agreement of the calculated structural data with the experiment demonstrates the important role of ionic binding and long-ranged Coulomb interactions in the HTSC's and leads already to a reasonable overall representation of the phonon dispersion within the RIM, as a first approximation of the insulating phase of these materials [37], see Fig. 1.

The various open symbols in this figure representing the experimental results for the insulating phase indicate the different irreducible representations characterizing the modes and the solid circles display the experimental values of the anomalous highest $\Delta_1$, and $\sum_1$ branch for the optimally doped metallic phase [3, 4, 41]. The strong renormalization of these OBSM modes upon doping, in particular of the $\Delta_1$ branch with the deep minimum ($\Delta_1 / 2$-anomaly or half-breathing mode, see Fig. 2) and, less pronounced, of the oxygen breathing mode, $O_B^X$, at the $X$ point, being the highest mode of the spectrum in the insulating phase, indicates a strong non-rigid electron-phonon coupling in the optimally doped metallic phase. A detailed study of these presumably generic phonon anomalies in the HTSC's will be given in Subsection 3.3. An understanding of the physical origin of these characteristic anomalies will bring light to the nature of the EPI and the electronic state in the cuprate superconductors.

Because in ionic materials like the HTSC's most of the electronic bonding charge is located at the ions a nonlocal coupling of the displaced ions in a phonon mode excites predominantly localized CF's in the shells of the $O_{xy}$ and $Cu$ ions in the $CuO$ plane. Such a physical picture is quantitatively realized in our approach to the density response, lattice dynamics and EPI of the HTSC's. The magnitudes of the CF's are dictated by the possibility to excite these CF's at the corresponding ions and in this context the degree of localization of the electrons is important. In particular the $Cu\ 3d$ states representing the most localized and correlated part of the electronic structure lead to large Coulomb-XC interactions ($\tilde{V}$ in Eq. (9)) which tend to suppress (phonon-induced) CF's. For an understanding of the EPI in the HTSC's the interplay of changes of the long-ranged Coulomb interactions, generated by the displaced ions favouring small $\bar{q}$ scattering, with corresponding screening effects introduced by the ionic CF's (and DF's) in a localized correlated electron system, largely controlled by the short-ranged part of the Coulomb interaction, is essential. In this context, the spin-degrees of freedom have indirectly influence on the CF's and vice versa the CF's influence the spin-fluctuations (SF's) via nonlocal EPI.

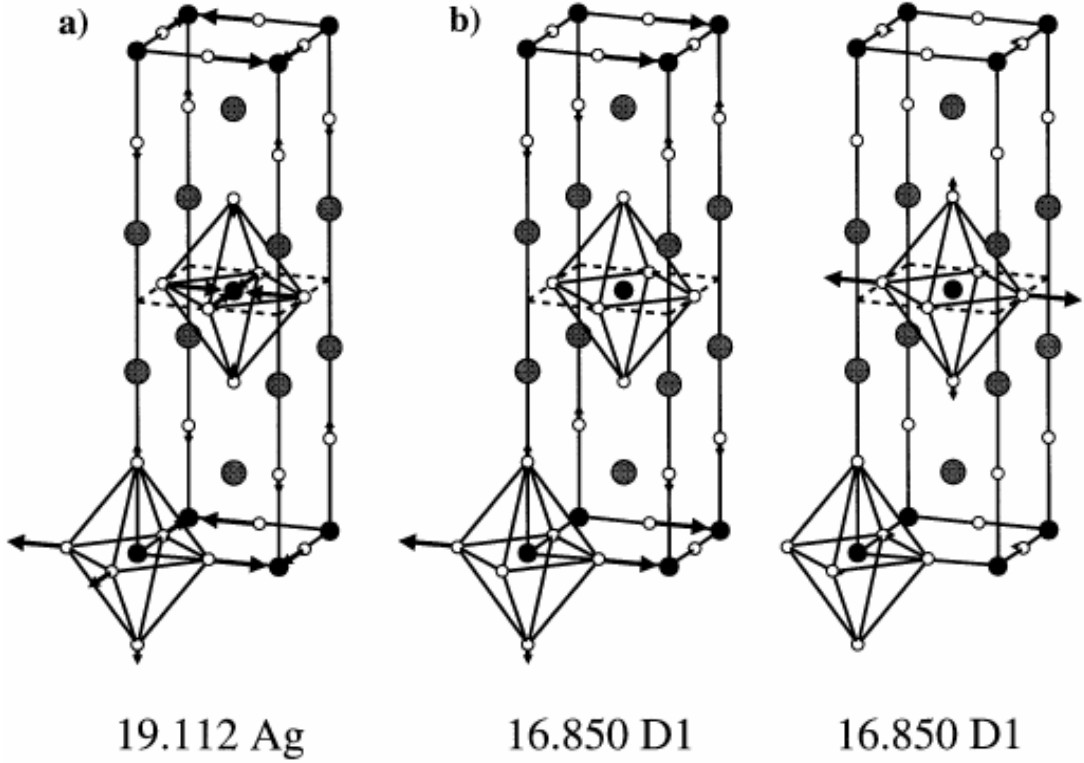

**Fig. 2** Displacement patterns of the high-frequency oxygen bond-streching modes of *La-Cu-O* in the metallic phase for the model corresponding to Fig. 4, (a) $O_B^x$ and (b) $\Delta_1 / 2$ minimum. In case of $\Delta_1 / 2$ the real part (left) and the imaginary part (right) are shown. Frequencies are in *THz*.

As can be expected from the above remarks and can be extracted from Fig. 1 the anomalous coupling and the corresponding phonon softening of the highest $\Delta_1$- and $\sum_1$ branch cannot be described by the RIM. This means that it must be related to the nonlocal, non-rigid EPI effects mediated by the second term in the dynamical matrix from Eq. (5). Interestingly, we observe in the RIM one partially unstable branch with the so called tilt mode at the *X* point. Freezing of this distortion points correctly to the experimentally observed structural phase transition from the high-temperature tetragonal (*HTT*) to the low-temperature orthorhombic (*LTO*) structure, see Fig. 3a. During this transition the *X* point of the tetragonal BZ is mapped onto the $\Gamma$ and *Z* point of the orthorhombic BZ and we find in our calculations for the low-temperature structure [42] two *stable* tilt modes with different frequencies whose displacement patterns and frequencies in *THz* are shown in Fig. 3b and 3c. Likewise as in the experiments [43] the frequency is higher for the tilt mode at $\Gamma$ than at $Z$ and the agreement with the experimental values (*2.5 THz* and *1.1 THz*) is reasonable, which shows that the structural phase transition is very well described by the ab initio RIM and, thus, essentially driven by the long-ranged ionic interactions in the material.

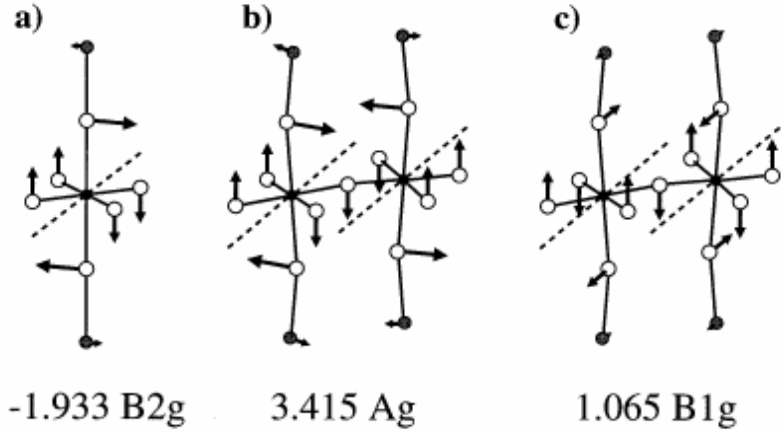

**Fig. 3** Displacement patterns of the tilt modes in *La-Cu-O* as calculated within the RIM. (a) tetragonal, *X* point; (b) orthorhombic, $\Gamma$ point and (c) orthorhombic, *Z* point. The broken line indicates the tilt axis for the structural phase transition *HTT* $\leftrightarrow$ *LTO.* The frequencies are in *THz* and the imaginary frequencies are represented as negative numbers.

While the high frequency of $O_B^X$ in Fig. 1 is favoured by the long-ranged Coulomb interaction between the ions, the so-called quadrupolar mode at the *X* point, which leads to a rhombic distortion of the oxygen squares around the *Cu* ions, is strongly destabilized by the long-ranged Coulomb forces [37]. Taking additionally the short-ranged interactions in the RIM into account we learn from Fig. 1 that the highest $\sum_4$ branch with the quadrupolar mode as its endpoint is already well described by the ionic reference system. For symmetry reasons CF's are not allowed in this mode. So we conclude that a quadrupolar deformation of the octahedra is very much promoted by the long-ranged Coulomb interaction. A phonon renormalization due to a *Jahn-Teller EPI* as proposed in [44] is unlikely and is at most of minor importance. Finally, we would like to comment on the modes propagating along the $\Lambda \sim (0,0,1)$ direction, i. e. along the *c*-axis. The doubly degenerate $\Lambda_3$ modes ( $E_g$ and $E_u$ at the $\Gamma$ -point) polarized parallel to the (*xy*)-plane are well described by the ionic reference system as can be seen from Fig. 1. On the other hand, in the case of the $\Lambda_1$ modes, which are polarized along the *c*-axis a considerable renormalization is required in order to get agreement with the experiments. As discussed in detail in [37], the main screening effect for these modes is due to anisotropic DF's with a dominating component along the ionic *c*-axis of the HTSC's.

Taking in addition to the calculations within the RIM CF's and anisotropic DF's into account we obtain a good overall agreement of the calculated phonon dispersion of *La-Cu-O* in the metallic as well as in the insulating phase, see Fig. 4 for the metallic and Fig. 5 for the insulating phase.

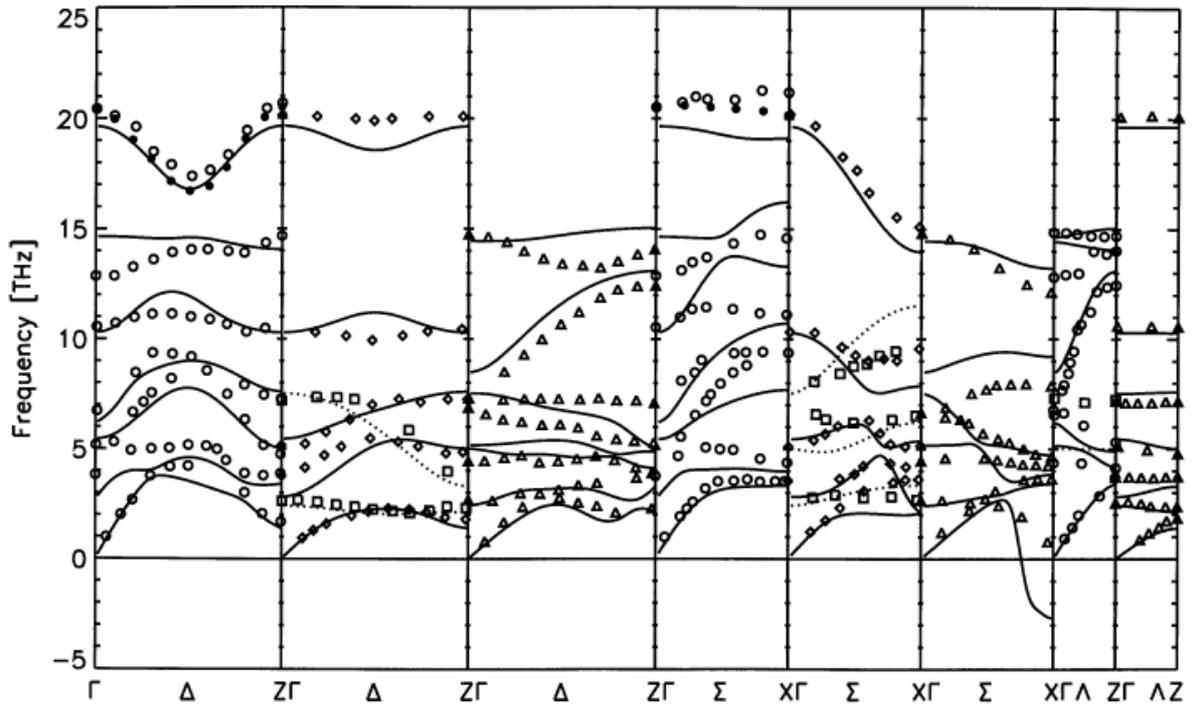

**Fig. 4** Calculated phonon dispersion for the metallic phase of tetragonal *La-Cu-O*. The experimental values for this phase $\left( La_{1.9}Sr_{0.1}CuO_4 \right)$, represented by various symbols as in Fig. 1, are taken from [3, 4, 41]. The notation for the ID's and the arrangement of the panels is the same as in Fig. 1.

These results demonstrate that a description of the electronic density response and the EPI by localized ionic CF's and anisotropic DF's is well suited for the HTSC's. A more detailed discussion of the results can be found in [37]. In the following we will address only some specific aspects of these calculations. Comparing Fig. 4 with Fig. 1 we recognize that in particular the anomalous softening of the high-frequency OBSM in the metallic phase is very well described by nonlocal EPI effects of CF-type. A thorough investigation of the anomalies is provided in Subsection 3.3. Anisotropic DF's practically play no role for this renormalization, as shown in part 3.3. In the insulating, undoped phase, Fig. 5, the anomalies are suppressed to a large extent because CF's are hard to excite due to the energy gap for the charges in the electronic spectrum. As already mentioned in the Introduction the model for the insulating phase has been designed to be consistent with the characteristic density response of an insulator, fulfilling sum rules in terms of the electronic polarizability $\Pi$ in the long-wave-length-limit $\vec{q} \rightarrow \vec{0}$ [34, 45].

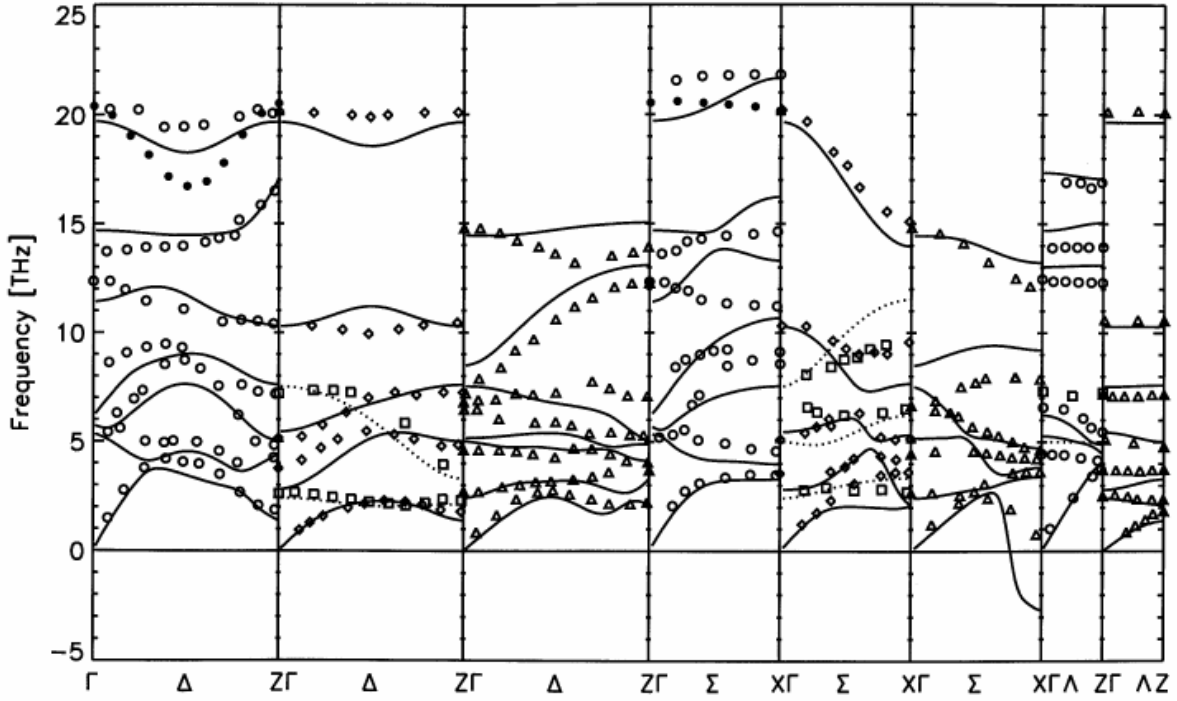

**Fig. 5** Calculated phonon dispersion for the insulating phase of tetragonal *La-Cu-O*. Experimental data and notation as in Fig. 1.

In the metallic phase the partial density of states (*PDOS*) of the quasiparticles at the Fermi level $Z_{\kappa}(\varepsilon_F)$ is related to the polarizability matrix at zero wave-vector according to

$$\sum_{\kappa'} \Pi_{\kappa\kappa'}(\vec{0}) = Z_{\kappa}(\varepsilon_F) \qquad , \qquad (19)$$

and the total density of states is defined as

$$Z(\varepsilon) = \sum_{\kappa} Z_{\kappa}(\varepsilon) \qquad . \qquad (20)$$

On the other hand, for an insulator we obtain the sum rules

$$\sum_{\kappa'} \Pi_{\kappa\kappa'}(\vec{q}) = O(q) \qquad (21)$$

and

$$\sum_{\kappa,\kappa'} \Pi_{\kappa\kappa'}(\vec{q}) = O(q^2) \qquad (22)$$

as $\vec{q} \to \vec{0}$ governing and restricting the electronic density response, in particular the CF's, in systems with an energy gap for charge excitations. This is different for the metallic case where charge excitations can appear as energetically low-lying excitations in the long-wave-length-limit. In the insulating phase spin-fluctuations are the low-lying excitations, but CF's are not completely suppressed only restricted and correlated according to the Eqs. (21), (22) in a special way. Equations (19, 20) and (21, 22), respectively, can be considered as an *orbital resolved* closed form to characterize the insulator-metal transition via the underdoped state in terms of the corresponding compressibility of the electronic system [34], as a primary tool to characterize the ground state of the electrons. Such orbital selective sum rules are particularly useful in the proximity to a Mott insulating phase where different from a band insulator the internal degrees of freedom, orbital and spin, still survive. For more details of the modelling, see [45].

Comparing Fig. 1 with Figs. 4 and 5 we find that now the axially polarized $\Lambda_1$ modes are well described. This renormalization effect as compared to the RIM is due to screening by anisotropic DF at the ions with a dominating $z$-component parallel to the $c$-axis [37].

A remarkable signature of the measured phonon dispersion in the metallic phase is the steeply dispersing $\Lambda_1$ branch, Fig. 4. As discussed in [31] a typical LDA-like bandstructure within the 31BM [46] taken as input for the calculation of the electronic polarizability $\Pi$ is not able to describe the strong anisotropy of the charge response in the $c$-direction seen in the experiments, while the charge response in the $CuO$ plane is quite well represented.

As a consequence the steep dispersion of the characteristic $\Lambda_1$ branch is largely underestimated and as shown in [37] a two-dimensional electronic structure for the calculation of the CF's results in a much better modelling of the charge response along the $c$-axis. In such a two-dimensional model only $Cu3d$, $4s$, $4p$ and $O_{xy}2p$ CF's in the $CuO$ plane are allowed and for the $\Lambda_1$ branch a good agreement with the experiment is obtained as can be extracted from Fig. 4. Moreover, in this calculation additionally off-site CF's at the centers of the planar oxygen sqares are introduced, simulating phonon induced distortions in the interstitial region. However, in such a situation of a two-dimensional or nearly two-dimensional electronic structure and thus a very weak interlayer coupling electron dynamics and phonon dynamics will be on the same time-scale along the $c$-axis and this calls for a non-adiabatic treatment with dynamical screening of the bare Coulomb interaction. For a general discussion of this problem and its relation to an apparent inconsistency between the current interpretation of the neutron scattering results for the $\Lambda_1$ branches in metallic samples as shown in Fig. 4, that look like those expected for an anisotropic metal in the adiabatic approximation, and the corresponding infrared-data being typical for an ionic insulator we refer to Subsection 3.4.

**Table 1** Calculated matrix elements of the macroscopic high-frequency dielectric tensor $\underline{\underline{\varepsilon}}_\infty$ and the static dielectric tensor $\underline{\underline{\varepsilon}}_0$ for tetragonal $La_2CuO_4$ together with an averaged value $\overline{\varepsilon} = \frac{1}{3}\left(\varepsilon_{xx} + \varepsilon_{yy} + \varepsilon_{zz}\right)$. RIM denotes the ab initio rigid ion model, CF includes additionally charge fluctuations, and CFD is the full model with CF's and anisotropic DF's.

| $La_2CuO_4$ | $\varepsilon_{zz}$ | | | $\varepsilon_{xx} = \varepsilon_{yy}$ | | | $\overline{\varepsilon}$ | | |
|---|---|---|---|---|---|---|---|---|---|
| | RIM | CF | CFD | RIM | CF | CFD | RIM | CF | CFD |
| $\underline{\underline{\varepsilon}}_\infty$ | 1 | 1 | 1.93 | 1 | 6.33 | 6.69 | 1 | 4.55 | 5.11 |
| $\underline{\underline{\varepsilon}}_0$ | 3.97 | 3.97 | 6.86 | 15.01 | 20.34 | 33.20 | 11.33 | 14.90 | 24.42 |

We conclude this Subsection by specifying some of the results we have obtained for the dielectric and infrared properties of *La-Cu-O* in the insulating state [37]. The calculations have been performed within the model leading to the phonon dispersion shown in Fig. 5. From the data listed in Table 1 we find that in this model, where CF's and DF's are allowed (model CFD), the matrix elements of the static dielectric tensor $\underline{\underline{\varepsilon}}_0$ are much larger than those of the macroscopic high-frequency dielectric tensor $\underline{\underline{\varepsilon}}_\infty$. This means that at low frequencies the optical phonons govern the dielectric behaviour. Comparing the calculated results for $\varepsilon_{\infty,xx}$ of the model CF, where only CF's in the *CuO* plane are admitted, with that of the model CFD, we see that the screening contribution of the electrons in the *CuO* plane is mainly due to the CF's. This is different than the situation in the classical ionic crystals, where the DF's are the most influential part [35]. On the other hand, along the *c*-axis screening is essentially caused by the DF's polarized predominantly along this axis expressing the ionic character between the layers. We also have calculated the transverse effective charges and the oscillator strength of the $A_{2u}$ and $E_u$ modes. The main message from these calculations is that the infrared response along the *c*-axis and perpendicular to it, each one of them, is dominated by just one strong phonon reflecting the ionic character of the material along the *c*-axis and the ionic layers, respectively. In case of the *c*-axis it is a "*ferroelectric-like*" $A_{2u}$ mode where the *Cu* and *La* cations vibrate coherently against the $O_{xy}$ and $O_Z$ anions generating a large dipole moment and in case of a polarization parallel to the *CuO* plane the strongest mode it is the $E_u$ mode with the lowest frequency where $La$ and $O_z$ are sliding in opposite direction to the $Cu$ and the planar oxygens.

### 3.2 Infrared- and Raman-active $c$-axis modes of $Bi$-based cuprate superconductors

In Ref [47] we have extended our studies within the present microscopic approach to the investigation of the $c$-axis lattice dynamics of the $Bi$-based cuprates. We have calculated the $c$-axis polarized infrared-(IR) and Raman-active modes for the single-layer $Bi_2Sr_2CuO_6$ ($Bi$-2201) and the bilayer $Bi_2Sr_2CaCu_2O_8$ ($Bi$-2212) superconductors. Our calculated results for the infrared-active $A_{2u}$ modes are in good agreement with new experiments obtained by ellipsometric measurements on single crystals [48]. The assignments of the $c$-axis $A_{2u}$-IR-phonons are discussed controversially in the literature [49, 50] and also as far as the Raman-active $A_{1g}$ modes are concerned different experimental groups have published rather different and contradictory assignments [51 – 53]. In particular the origin of the high-frequency vibrations around 460 and 630 cm$^{-1}$ in the Raman spectra of $Bi$-2212 is controversial and there is a strong dispute concerning the question of which mode can be attributed to the oxygens $O2$ located in the $SrO$ plane and which one to the oxygens $O3$ in the $BiO$ plane. In our calculations [47] he have provided a detailed assignment of all of the $A_{2u}$ and $A_{1g}$ modes in terms of the components of the corresponding eigenvectors. Our results concerning the problem with the Raman modes clearly assign the 630 cm$^{-1}$ phonon to the vibrations of the apical oxygens $O2$ located in the $SrO$ layers. Theoretical calculations of Raman- and infrared active modes have been presented within the empirical shell model [48, 54] and a force constant model [55], respectively. Very recently, calculations on the Raman modes of $Bi$-2212 within DFT-LDA have been performed [106] where structure optimization effects on the electronic and vibrational properties are studied which might also be useful for mode assignments in the experiments. The results in this work support the mode assignment as found in our calculations for the vibration at 630 cm$^{-1}$.In the following we present an excerpt of our calculations for the bilayer case and list the displacement patterns and frequencies of the infrared- and Raman-active modes. For a detailed discussion of mode assignment, a comparison with experiment and of the particular screening effects provided by anisotropic DF's and CF's for the renormalization of the phonon modes we refer to [47]. Here we would like to emphasize a particular interesting aspect of the calculations. A theoretically and experimentally important question, which has been attacked in [47] from the phononic side, is the possibility of a metallic character of the *BiO* layers. According to first-principles calculations of the electronic structure [18, 56] the *BiO* layers contribute to bands of *Bi p* and *O p* character which dip below the Femi energy in a small part of the BZ. These bands may act as a metallic charge reservoir which provides electron pockets that allow for an increase in the number of holes in the *CuO* bands, promoting (super-) conductivity in the *CuO* layers. In the real material, however, the matter is far more complicated and it is currently not possible to account in the first-principles calculations for the incommensurate complex superstructures in the *BiO* planes observed by diffraction experiments. The presence of the superstructure may also be a reason which greatly complicates the comparison with the experimental results obtained from photoemission, where the *BiO*-derived pockets predicted from theory, cannot yet be resolved [57].

**Table 2** Comparison of the calculated eigenfrequencies of the $A_{2u}$ modes of *Bi-2212* in units of cm$^{-1}$ for different models as explained in the text with experimental results.

| Bi-2212 | $A_{2u}(1)$ | $A_{2u}(2)$ | $A_{2u}(3)$ | $A_{2u}(4)$ | $A_{2u}(5)$ | $A_{2u}(6)$ |
|---|---|---|---|---|---|---|
| SM [48] | 93 | 195 | 304 | 343 | 403 | 643 |
| SM [54] | 137 | 169 | 277 | 334 | 487 | 514 |
| FM [55] | 164 | 225 | 281 | 442 | 467 | 617 |
| RIM | 79 | 150 | 285 | 360 | 595 | 700 |
| C | 74 | 135 | 195 | 268 | 462 | 618 |
| CB | 73 | 134 | 195 | 267 | 354 | 604 |
| Exp. [48] | 97 | 168 | 210 | 304 | 358 | 583 |

**Table 3** Comparison of the calculated eigenfrequencies of the $A_{1g}$ modes of *Bi-2212* in units of cm$^{-1}$ for different models as explained in the text with experimental results.

| Bi-2212 | $A_{1g}(1)$ | $A_{1g}(2)$ | $A_{1g}(3)$ | $A_{1g}(4)$ | $A_{1g}(5)$ | $A_{1g}(6)$ |
|---|---|---|---|---|---|---|
| SM [48] | 112 | 149 | 208 | 370 | 461 | 630 |
| SM [54] | 87 | 164 | 182 | 387 | 494 | 517 |
| FM [55] | 80 | 123 | 170 | 316 | 459 | 616 |
| RIM | 125 | 157 | 196 | 493 | 598 | 730 |
| C | 115 | 147 | 178 | 399 | 508 | 656 |
| CB | 98 | 144 | 162 | 382 | 451 | 608 |
| Exp. [51] | 59 | 117 | 145 | 409 | 463 | 627 |

We have investigated the effect of the *BiO* layers on the phonons by calculating the phonon modes within two different models. One model, *CB* in Tables 2 and 3, allows for metallic CF's in both the *CuO* and *BiO* layer and a second model, *C*, where the CF's in the *BiO* layers are suppressed and only CF's in the *CuO* layers are admitted. Additionally, we have included in the tables the results as obtained for the RIM and the empirical models. A comparison of the data as obtained with the RIM with those of models *C* or *CB*, respectively, makes visible the screening effects provided by the nonlocal EPI via CF's and DF's. Comparing the calculated results for the infrared active $A_{2u}$ (5) mode in Table 2 for model *C* and model *CB*, respectively, we find a very strong renormalization of this mode induced by the CF's in the *BiO* layers resulting in a frequency which is in excellent agreement with the experiments. Note that the empirical models are unable to provide a good fit for this mode. The calculated frequency in model *C* is *462* cm$^{-1}$ in model *CB 354* cm$^{-1}$ and in the experiment *358* cm$^{-1}$. From an investigation of the phonon-induced CF's according to Eq. (12) for this mode in model *C*, only a charge transfer (CT) between the *CuO* layers is generated, while in model *CB* we obtain, additionally to the CT between the *CuO* layers, a strong CT between the *BiO* layers. The additional metallic CF's at *Bi* and *O3* lead to the correct energy relaxation and to a very good agreement with the experiment. In Fig. 6 the corresponding displacement patterns of the $A_{2u}$ modes are shown.

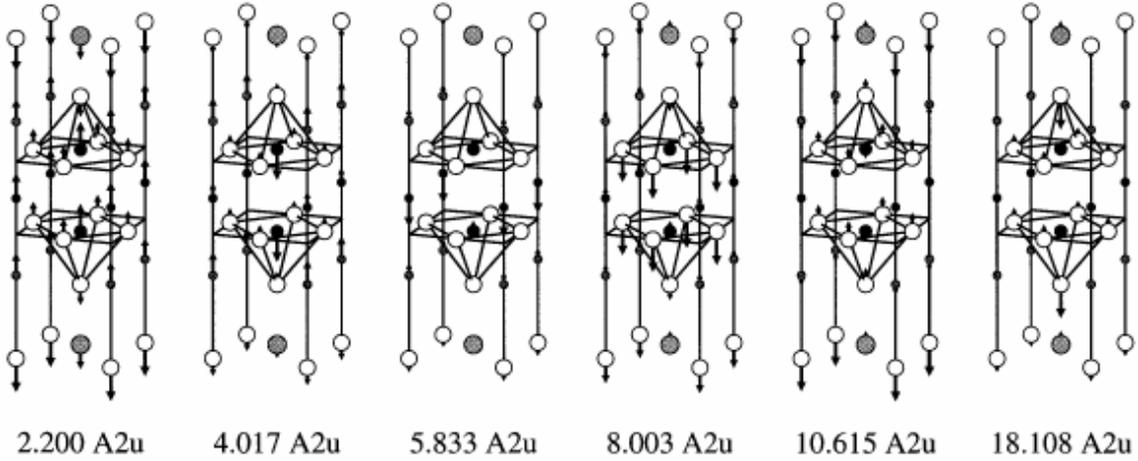

2.200 A2u    4.017 A2u    5.833 A2u    8.003 A2u    10.615 A2u    18.108 A2u

**Fig. 6** Displacement patterns of the six infrared-active $A_{2u}$ symmetry modes in $Bi_2Sr_2CaCu_2O_8$ as calculated from model *CB*. The frequency below each pattern is given in *THz*.

Similar strong phonon renormalization effects introduced by the CF's in the *BiO* layers are found for the Raman-active modes $A_{1g}$ (5) and $A_{1g}$ (6), see Table 3 and Fig. 7, and also in the single layer materials for both type of modes [47].

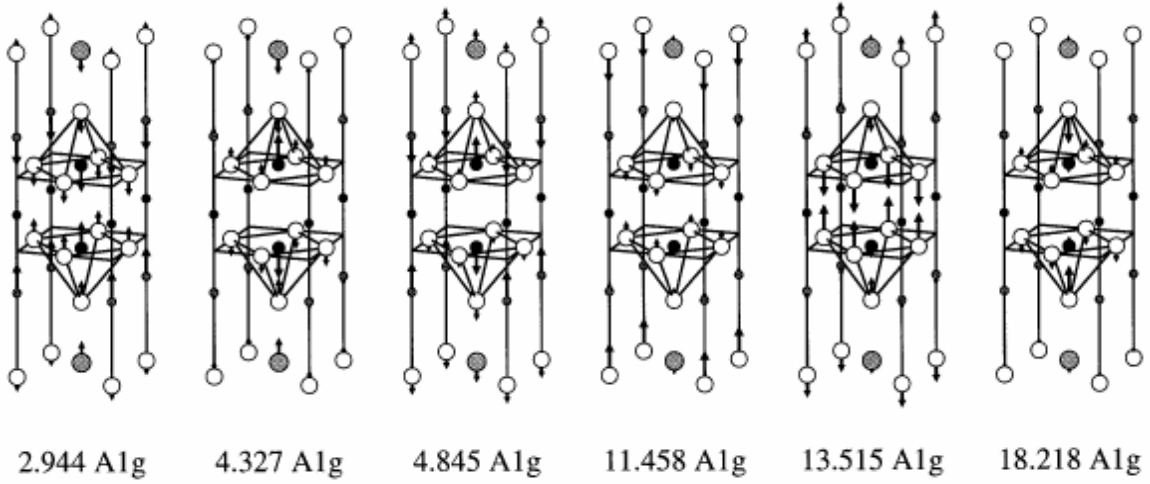

2.944 A1g    4.327 A1g    4.845 A1g    11.458 A1g    13.515 A1g    18.218 A1g

**Fig. 7** Displacement patterns of the six Raman-active $A_{1g}$ symmetry modes in $Bi_2Sr_2CaCu_2O_8$ as calculated from model *CB*. The frequency below each pattern is given in *THz*.

Thus, from the phononic side evidence of the metallic behaviour of the *BiO* layers is provided. This speaks in favour of the opinion that the *BiO* planes may play the role of an electron attracting reservoir and therefore may dope the *CuO* layers with holes. Finally, we detect, similarly as in *La-Cu-O*, in both *Bi*-compounds "*ferroelectric-like*" modes where the cations vibrate coherently against the anions and as a consequence the dipole moments generated by the motion of the ions add constructively to a large value. These modes, see the $A_{2u}$ *(4)* mode in *Bi-2212* in Fig. 6, are expected to contribute significantly to the low-frequency optical *c*-axis response.

### 3.3 Generic phonon anomalies and modelling of the electronic state in the HTSC's

Despite an enormous experimental and theoretical effort the electronic state and the mechanism of superconductivity in the HTSC's is unclear. There is not even consensus as to the simplest "effective Hamiltonian" needed to account qualitatively for superconductivity and the unusual normal state properties in the cuprates. Quite general the physics results from an interplay of lattice-, charge- (orbital-) and spin degrees of freedom. The interaction of two of them, lattice and charge (orbital), will be studied explicitly in this section by means of the generic high-frequency OBSM in *La-Cu-O* showing an anomalous softening upon doping in the metallic phase (Fig. 1) which indicates a strong nonlocal EPI. How the spin-degrees of freedom come into play is also illustrated qualitatively by means of these modes. In particular we investigate the half-breathing mode ($\Delta_1 / 2$ mode) where the renormalization is strongest and the so called planar breathing mode $\left(O_B^X\right)$. The displacement pattern for each of these modes is shown in Fig. 2. In $O_B^X$ essentially the planar oxygens vibrate against (or away from) the silent copper and in $\Delta_1/2$ the $O_y$ and the $Cu$ are silent in case the $O_x$ are vibrating and vice versa. Already in Ref. [34] we have pointed out on the basis of our microscopic modelling of the density response in the HTSC's that the OBSM mix strongly with the CF's leading to the observed anomalous softening. Moreover, an explanation of the experimentally observed *anisotropy* in softening, i. e. that the phonon renormalization is stronger in the $\Delta \sim (1,0,0)$ direction than in the $\Sigma \sim (1,1,0)$ direction (Fig. 1) has been provided. For symmetry reasons there are only CF's at the silent $Cu$ allowed in $O_B^X$ and not on the moving $O_{xy}$ ions. On the other hand in the $\Delta_1/2$ mode in addition to the CF's at the $Cu$ ion also CF's can be generated at the silent oxygens, even more easily, because of the smaller on-site Coulomb interaction for *O 2p* states compared to the strongly localized *Cu 3d* states. This leads to an additional source of softening in case of $\Delta_1/2$ as compared to $O_B^X$. In the following publications [45, 58] the phonon anomalies have been reinvestigated within a more realistic ab-initio RIM as reference system and an improved modelling of the electronic polarizability. From theses calculations results have been obtained which are in a good quantitative agreement with the experiments. Moreover, in [45] the experimentally observed softening of the OBSM during the insulator-metal transition via the underdoped phase has been studied. Modes outside the non-adiabatic region, see Subsection 3.4, can be well described in the adiabatic approximation. Only for $O_B^X$ a small non-adiabatic correction of about *0.1 THz* has been found which can be attributed to Fermi surface nesting [46]. In Fig. 8 a comparison of the calculated results for the anomalies with the experiments for the optimal doped metallic and the insulating phase is shown. In these calculations a two-dimensional electronic structure for the CF's in the *CuO* plane and anisotropic DF's as for the calculations of the full phonon dispersion in section 3.1 are taken into account. The effect of the latter on the softening has not been calculated in [45, 58]. From the results given in Fig. 8 we see that the additional screening by DF's is not essential for these modes. It can be related to anisotropic DF's along the *c*-axis of the $O_z$ and *La* ions in the ionic layers.

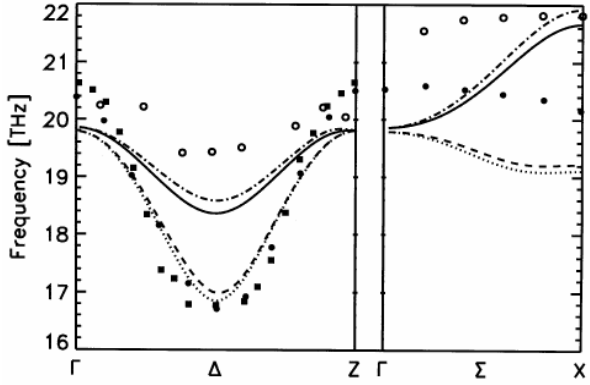

**Fig. 8** Calculated phonon dispersion of the highest $\Delta_1$ and $\Sigma_1$ branch (phonon anomalies) for different models: $-\cdot-$: insulator CF's only, $-$: insulator CF's and anisotropic DF's; $---$: metal CF's only; $\cdots$: metal CF's and anisotropic DF's. The various symbols indicate the experimental results for the insulating phase (o) and metallic $La_{1.85}Sr_{0.15}CuO_4$ (●,■) [3, 4].

For the underdoped phase of the (*p-doped*) HTSC's, where an accurate description of the low-energy excitations remains a theoretical challenge and where the question how the metallic charge conduction emerges from the doped Mott insulator is poorly understood, we have proposed as a bridge a model for the electronic density response (pseudo-gap model [45]) to describe this localization–delocalization transition in terms of the compressibility, which needs the orbital freedom of *Cu 3d* and *O 2p* states and assumes that the single-particle partial density of states at the Fermi level is suppressed for the strongly correlated localized *Cu 3d* states because of the cost in energy from hopping of the charge carriers to the *Cu*-sites, (insulator-like, *local orbital selective,* incompressible charge response; Eq. (21)), but not so for the more delocalized *O 2p* orbitals where the holes predominantly are injected to in *p-type* cuprates (metallic, *local orbital selective,* compressible charge response with a *renormalized PDOS*; Eq. (19)). So we have partially incompressible regions with local moments of spin, however, the total compressibility is never zero in this phase, a real space organization for the low lying charge exitations is achieved and a charge transport is enabled without strong hindrance by mutual Coulomb repulsion. On the other hand, the incompressible regions compete with overall metallic behaviour and with superconductivity. Moreover, a loss in the density of states of the quasiparticles in the *CuO* plane at the Fermi level is consistent with such a model and a metallic mobility behaviour is possible only along the oxygen network in the *CuO* plane but blocked at the *Cu* sites which qualitatively should lead to a reduced screening, to the high electrical resistivity and the small carrier number for the underdoped material experimentally observed. Note, however, that a lowering of the energy would be allowed by virtually hopping to the *Cu* states which will couple the charge transport to the remaining antiferromagnetic correlations in the underdoped material. In this way, both, charge- and spin-degrees of freedom contribute to the gap. Likewise as in the insulating phase the more delocalized higher energy *Cu 4s (4p)* states are not admitted to be occupied in the model for the underdoped phase in contrast to the case at higher doping, see below. The partial "*correlation-gap*" provided by the *Cu 3d* orbitals in our modelling leads according to Eq. (21) to strongly reduced insulator-like CF's at the *Cu*. The calculated phonon dispersion within such a model is in good agreement with the measured dispersion for the underdoped phase of *La-Cu-O* [45], see also Fig. 9, where the effect of the insulator-metal transition via the underdoped regime on the phonon anomalies is shown. The strong increase of the frequency of the OBSM with reduced doping also reflects that screening of the bare long-ranged Coulomb interaction which is responsible for the high values of the modes in the insulating state (or the RIM, respectively) becomes less effective approaching the metal-insulator transition (insulator-like charge response of the correlated *Cu 3d* states, *correlation gap*).

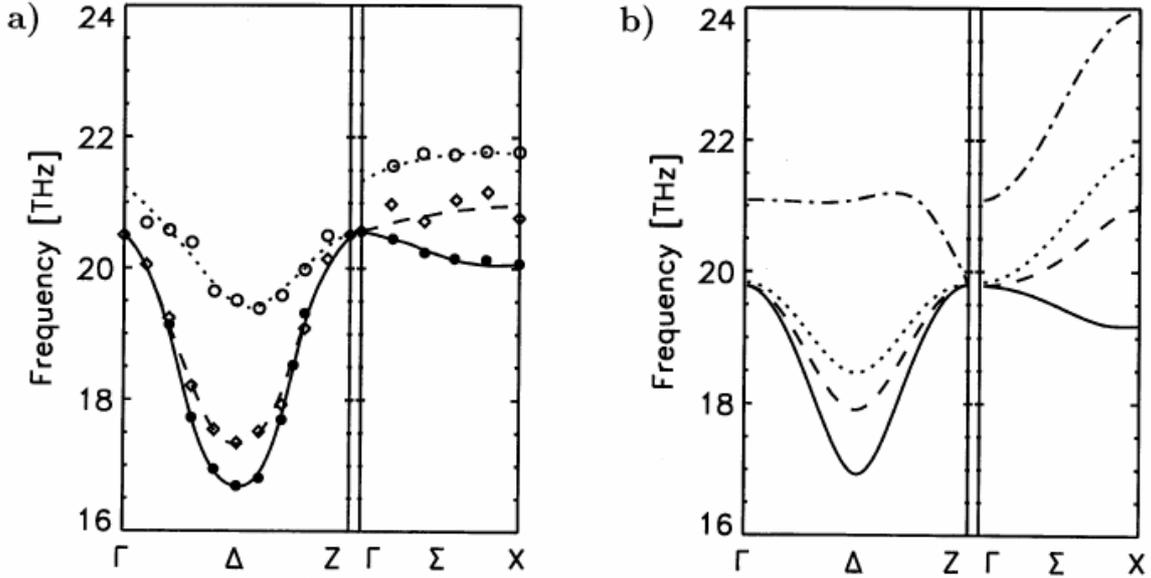

**Fig. 9** (a) Experimental results for the highest $\Delta_1$ and $\Sigma_1$ branch of $La_{2-x}Sr_xCuO_4$ according to [3, 4, 41]. The symbols o represent the insulating phase $(x = 0)$, • corresponds to the optimally doped metallic phase $(x = 0.15)$ and ◊ belongs to the underdoped phase $(x = 0.1)$. The lines are a guide to the eye. (b) Calculated results for the phonon branches shown in (a) according to Ref. [45]. The dotted curve gives the results for the insulating phase, the dashed curve for the underdoped phase, and the full curve for the optimally doped phase. For comparison, the results for the ab initio rigid-ion model without any CF's are also displayed (dashed-dotted curve), which allows to extract the effect of the nonlocal EPI.

Our microscopic description of the underdoped phase of the HTSC's in form of such a textured "two-component" electronic structure establishes a bridge between the antiferromagnetic insulator state of the undoped material and the more conventional metallic state at higher doping. Note, that by applying Eqs. (19, 21) both, information about the energy spectrum as well as the wavefunctions is introduced into the modelling. The model simulates a novel metallic state where the metallic texture of the "*O 2p-skeleton*" of the oxygen sublattices in the *CuO* plane is penetrated with an insulator-like "*Cu 3d-sublattice*". In this way the metallic charge transport along the *O 2p network* is in principle implicitly related to the antiferromagnetic correlations of the *Cu* spins not explicitly considered in our modelling. So in principle both is possible, low-energy charge excitations on the *O 2p network* and separated from these charge excitations spin excitations on the *Cu sublattice* accompanying the charge transport. It is worthwhile to mention that in such a desciption the charge transport in the underdoped cuprates is not caused by the motion of ordinary quasiparticles (QP) in a two – dimensional electronic system but is restricted to a textured subspace of the two-dimensional space defined by the oxygen network. Moreover, the system is not an insulator (or semiconductor) with a well defined gap. The oxygen derived charge states of the skeleton may be interpreted as in-gap states as compared with the undoped case with a gap and spectral weight can be expected to be transferred with doping. From an energetic point of view such a microscopic model seems to be favourable because via the metallic oxygen-skeleton there is a gain in kinetic energy for the charge carriers by delocalization avoiding simultaneously the large Coulomb-repulsion at the *Cu* while due to the insulating copper-sublattice the antiferromagnetic correlations of the parent insulator are partly preserved (depending on the degree of doping; note that the pure antiferromagnetic insulator tries to maximize the exchange energy). Moreover, the kinetic energy cost introduced by the opening of the "*correlation gap*" can be restored.

In case of sufficiently strong residual antiferromagnetic correlations at low doping, the coherent propagation of a *pair* of two holes might be favoured by the kinetic energy as discussed for example in [59]. In such a situation the "*correlation gap*" at *Cu* would favour pairing already in the normal state. Such a precursor pair for superconductivity would also result in a gap in the spectrum ("*pseudogap*") which, however, differs from the "*correlation gap*" which itself could be interpreted as a precursor gap of the Mott insulator. Nevertheless, both types of gaps would be linked by the correlations introduced through the localized *Cu 3d* states. Note, however, if the possibility of precursor pairs of some kind in the normal state is theoretically explored, the strong nonlocal EPI with the corresponding changes in the potential an electron feels (Eq. (15)) and the CF's themselves found in our calculations should be considered besides the electron-electron interaction effects that lead to spin-correlations in a many-body treatment of such preformed pair states and the resulting gap in the normal as well as in the superconducting state. The same holds true for the charge excitations on the oxygen network. In the experiments also a gap, in general called pseudogap, appears in the underdoped side of the phase diagram of the HTSC's which weakens as optimal doping is approached, see [60] for a review.

Quite recently an extensive ARPES study of various lightly doped cuprates [61] has found a finite gap over the entire Brillouin zone, including the nodal line, which closes with increasing doping. In our model picture this would mean that the charge response at the oxygen network additionally becomes insulator-like with a gap at very low doping. However, this is not the result of a further "*correlation gap*" as in the case of the *Cu* sublattice. It is likely that the ionic nature of the HTSC's with the long-ranged Coulomb interactions, only poorly screened at very low doping, is important for the additional loss in density of states at the Fermi level. One might speculate as is done in [61] that at a sufficiently low doping rate the strongly reduced screening of the long-ranged Coulomb interaction of the disorder potential of the dopants is the reason for charge localization and it is known in this context that a synthesis of disorder (introduced e. g. by doping) with long-ranged Coulomb interactions can create a "*Coulomb gap*" for single-particle excitations at the Fermi level [62].

It should be noted, that the various inhomogeneous charge and spin ordered configurations discussed in the literature as possible low energy states can formally be considered as certain "defect structures" of our model if the spin-degrees of freedom are explicitly considered. A possible defect structure which could arise from our model would be a texture where hole-rich oxygen centered charge regions with possibly more delocalized *Cu*-related states at the neighbouring *Cu* ions (e. g. due to an enhanced *Cu 4s* occupation and a reduced *Cu 3d* occupation, compare the discussion below) and consequently a reduced antiferromagnetic interaction alternate with regions with localized *Cu* states (reduced *4s* occupation) and simultaneously a strong antiferromagnetic interaction. This kind of self-organization of the charge carriers might possibly be realized at the threshold of the insulator metal transition at very low doping. Moreover, such a hypothetical system may gain kinetic energy with larger doping via an increased population of the *Cu 4s* states of the metallic regions leading to a hybridization between the latter.

In a concluding remark the different situation for *n-doped* cuprates is shortly addressed. In this context an important *electron-hole asymmetry* becomes visible in our modelling, i. e. the different character of hole and electron carriers. While the former are doped primarily to the oxygen sites in the *CuO* plane with *O 2p* character the latter will preferentially go to the *Cu* site with *Cu 3d-4s* character. Thus, within our modelling of the electronic density response in contrast to the hole-doped case the underdoped phase of the *n-type* material should be modelled in terms of a metallic charge response according to Eq. (19) at the *Cu* site with an increasing *Cu 4s* occupation upon doping (this leads to a smaller effective on-site U parameter for *Cu* than in the hole-doped case in a model approach using a reduced set of degrees of freedom because of better screening) and a localized insulator-like response additionally at the $O_{xy}$ sublattices according to Eq. (21). The latter selfconsistently may stabilize antiferromagnetic order by localization which is much more robust in the electron-doped material and the superconducting range is much narrower. Then, in the optimally doped state a crossover to a metallic charge response (Eq. (19)), at the formerly "insulating" $O_{xy}$ sites (and the *Cu 3d* orbitals) would be appropriate (see the discussion for the *p-doped* case below). From such a modelling of the electronic state it becomes obvious that besides the spin response also the EPI can expected to be different in hole-doped and electron-doped materials reflecting an *electron-hole asymmetry*. So far we have not investigated quantitatively these important connections.

According to our calculations the basic physical reason for the strong phonon renormalization in the OBSM are nonlocal EPI effects in form of CF's localized on the electronic shells of the ions in the *CuO* planes. These coupling effects are transferred to the phonon dispersion by the second term in the expression of the dynamical matrix in Eq. (5). This particular type of screening in terms of ionic CF's dominates in the HTSC's because most of the electronic bonding charge is localized at the ions, completely different from the homogeneous electron gas picture of conventional metals and superconductors where local EPI effects dominate. Our theoretical method is particularly well adapted to a description of the density response of the strongly localized *Cu 3d* states in direct space for which correlation effects are important and that retain their atomic character in the HTSC's to a large extent. Following Eqs. (3, 7, 9, 11) the approach is well suited to describe materials made up of electrons that are neither fully itinerant nor fully localized because we take via $\Pi$ (kinetic quasiparticle part) the momentum-space and via $\tilde{V}$ (correlation features etc.) the real-space picture being particularly important for localized electronic states into account allowing, moreover, to incorporate the material-specific details into the calculations.

The blocking of the metallic charge response at the *Cu* sites in the underdoped phase of the *p-doped* cuprates is lifted in our modelling of the optimally doped metallic phase by the fulfilment of the sum rule in Eq. (19), metallic behaviour is now additionally possible via the *Cu-O* links. In a $\vec{k}$-space picture a large Fermi surface (FS) as found in LDA-calculations can now develop from a gapped FS of the underdoped phase, consistent with the fact that the calculated electronic polarizability $\Pi$ of Eq. (11) is derived from a LDA bandstructure in the optimally doped case. Thus we have a crossover in the electronic properties corresponding to a qualitative change of the groundstate. As has been demonstrated in [58] the amplification of softening of the OBSM in the optimally doped metallic phase of *La-Cu-O* can be explained by the growing importance of the CF's in the more extended (delocalized) orbitals of the electronic structure, besides *O 2p* states mainly the higher energy state *Cu 4s* and less important also *Cu 4p* is concerned. Increasing the population of the latter states makes the system and the charge transport more Fermi-liquid-like and also increases the three-dimensional nature of the material. In this way an additional gain in kinetic energy is obtained mainly due to delocalization via *Cu 4s*, without paying a total penalty for the exchange interaction, related to the remaining antiferromagnetic correlations mediated by the localized *Cu 3d* states. Obviously, such a situation seems to be favourable also for superconductivity. Specifically, the additional renormalization is related to the contribution of mainly the *Cu 4s* and to a smaller part the *Cu 4p* orbitals despite their relatively weak occupation, see subsection 3.1. This is consistent with the fact that the large on-site repulsion $U_{dd}$ of the strongly localized *Cu 3d* orbital leads in tendency to a suppression of low-energy CF's according to Eq. (12) via the quantity $\vec{X}$ from Eq. (14) or $C(\vec{V})$, respectively, while, on the other hand, the more extended *Cu 4s* and *Cu 4p* orbitals, with a strongly reduced on-site repulsion $U_{ss}$ and $U_{pp}$, allow for an increase of delocalized CF's at the Cu ion and an enhancement of the itinerant character of the electronic structure. Note, that also the calculated on-site parameters at *Cu*, $U_{sd}$, $U_{pd}$, $U_{sp}$, are much smaller than $U_{dd}$. Quite generally the cost in potential energy because of the Coulomb repulsion arising from the hopping of the electrons must be exceeded by the gain in kinetic energy from delocalization and this is achieved much easier for the *Cu 4s, 4p* and the *O 2p* states. Further experimental evidence other than phonons for the growing importance of the delocalized states in case of the optimally doped probe comes from optical conductivity measurements [63] that demonstrate that already in the optimally doped regime with highest $T_c$ an itinerant carrier contribution has developed. It is tempting to relate the Drude component in the optical response to the delocalized, ungapped states with low-energy CF's and the broad band at higher energy to the localized *Cu* states with higher energy CF's, where upon doping also the *Cu*-states start to delocalize most likely mainly via *4s* and contribute besides the *O 2p* states to the itinerant motion of the charge carriers. Thus, from our findings for the phonon softening we see that a critical degree of delocalization in particular of the *Cu*-related states introduced mainly via a population of the *Cu 4s* orbital of higher energy (and a corresponding reduction of the occupation of the *O 2p* and *Cu 3d* states), which also determines in a specific way the range of electron hopping and hybridization with the other states in the neighbourhood, is important for the strength of the anomalies in the nonlocal coupling regime at optimal doping and simultaneously as seen from the experiments for superconductivity and the observed critical temperature $T_c$. From our calculations a broad orbital freedom and the full three-dimensional Coulomb interaction is required for a realistic description of the phonon anomalies and even small changes in localization of these states are an important tuning parameter.

The enhanced screening effect of the delocalized *Cu* states, *4s (4p)* can also be discussed at least qualitatively in a simpler model with a reduced set of electronic degrees of freedom like in [34]. In this work CF's are only allowed for the *Cu 3d* and the $O_{xy}$ *2p* states of the ions in the *CuO* planes. The contribution of the more extended states is simulated in this model by renormalizing (screening) the on-site Coulomb-XC interactions. Using such reasoning, the anomalous softening for $\Delta_1 / 2$ and $O_B^x$ now seen in the experiments already has been predicted in [34] and related to the changes of the strong repulsive Coulomb interaction at the *Cu* ion as an important physical parameter of the cuprates.

For high doping in the overdoped phase the character of the electronic state changes again and it can be expected that the importance of the delocalized states in the CuO plane is further increased. From the optical conductivity spectra it can be seen that the Drude-type band rapidly grows with doping [63] indicative of an electronic state of much enhanced itinerant character. Doping holes into the *CuO* plane (besides for the *O 2p* states) also creates space for a delocalization of the *Cu*-related states. Thus, the occupation of the delocalized component of the *Cu* states, i. e. the *Cu 4s (4p)* states, will be further increased as compared to the optimally doped case. Such a scenario will appear in the normal state of the HTSC's if the gain in kinetic energy due to delocalization and hybridization with the orbitals in the neighbourhood specific for the corresponding material outweighs the loss in binding energy related to the decreased localization of the electrons and the loss in exchange energy, which is an important contribution in the underdoped phase and in the optimally doped phase, too, as discussed above. In particular, we have a competition between the exchange energy favoured by localization and the kinetic energy favoured by delocalization, where the latter seems to win in the heavily doped cuprates.

The delocalization will also enhance the single-particle contribution to the interlayer coupling which leaves characteristic fingerprints in the phonon-plasmon scenario as discussed in Subsection 3.4. On the other hand, a suppressed delocalization in the underdoped phase will suppress the coherent component of the *c*-axis charge transport.

As far as the character of the EPI is concerned one can say that increasing doping changes the latter from a strongly nonlocal interaction in the underdoped and optimally doped regime to a more local one in the overdoped case. Such a qualitative change of the EPI upon doping should also be reflected in the self-energy of the quasiparticles (higher scattering rate in the underdoped and optimally doped case) and detectable e. g. in modern photoemission experiments with high resolution [64]. In the superconducting state the EPI and its effect on the self-energy is further modified because low-energy phonon scattering is blocked in an anisotropic way according to the symmetry of the gap. In total, there is another possible doping dependent boson scattering channel which should be considered for the interpretation of the renormalization of the quasi particle self-energy together with the magnetic excitations, commonly invoked, and the CF's.

Also the low-energy plasmons (intraband- and interband plasmons along the $c$-axis) studied in Part 3.4 accounting for the three-dimensionality of the actual material are a possible source for a doping dependent renormalization of the self-energy and in particular, for a suppression of the quasiparticle weight [65] (e. g. for model M1 in Part 3.4 the coherent quasiparticle transport along the $c$-axis is completely switched off and an acoustic plasmon evolves). Finally, material dependent changes of the FS, see below, and the doping dependence of the FS itself should not be ignored, in particular at $(\pi, 0, 0)$, see Fig. 10, when investigating electronic self-energy effects and interpreting the latter in the experiments.

In summary, the enhanced delocalization in the overdoped phase should lead to a gain in kinetic single-particle energy in the normal state, a suppression of the antiferromagnetic correlations. and simultaneously to a weakening of the strongly nonlocal character of the EPI favoured in particular by the localized $Cu\ 3d$ states. In this way the constructive contributions via combined lattice-, localized charge- and spin-degrees of freedom brought about by correlation effects to pairing in the HTSC's in the optimally doped phase are weakened. In other words, on the route towards the overdoped state upon doping conventional metallic screening is strengthened by the growing importance of the delocalized CF's and many body interactions related to localization are reduced. As a consequence also the resonant magnetic excitations observed in the superconducting state are expected to disappear gradually. Simultaneously, only the more local EPI and CF's with an enhanced delocalized component essentially remain as a source for pairing while the contribution of the spin-degrees of freedom dies out with the vanishing of the antiferromagnetic correlations. For the behaviour of the phonon anomalies this means, that on account of our calculated results obtained for the optimally doped phase as a function of delocalization in the overdoped material a further softening can be expected because of an increased population of the delocalized *Cu 4s (4p)* states. In order to obtain optimal conditions for superconductivity a delicate balance between the occupation of localized and delocalized states as found in our calculations is required. Having recognized delocalization upon doping as an important tuning parameter for the phonon anomalies, it is quite interesting to see, what may be expected from these findings for the doping dependence of the QP band-structure, in particular for the generic "flat-band" around $(\pi,0,0)$ in the cuprates. This band is below the Fermi energy $E_F$ for small doping, moves upwards monotonically with hole doping towards $E_F$ and crosses the Fermi level in the overdoped region, thereby the flatness is somewhat reduced and dispersion is generated [66, 67]. Leaving possible effects provided by correlation aside, which may contribute to the flatness of the band, part of this behaviour could be interpreted as follows. Both, the upward shift and the deformation of the band upon doping could be qualitatively comprehended because the delocalization related to the *Cu 4s* states as found in our calculations for the OBSM should lead via the increased *4s* occupation to an upward shift of the band center accompanied by a certain downward pushing through the coupling of the *4s* orbital to the *pdσ* orbitals, in addition the interlayer coupling can be expected to increase.

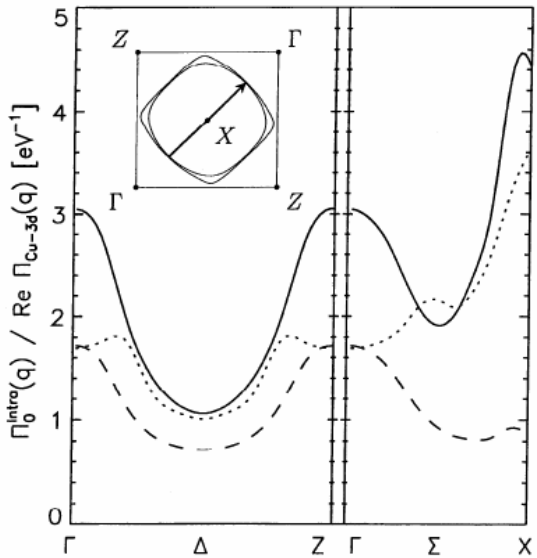

**Fig. 10** Intraband contribution of the scalar electronic polarizability $\Pi_o\left(\vec{q}\right)$, according to Eq. (23) for *La-Cu-O* within the 31 band model for the Fermi energy $E_F = 0$ (dotted curve) and $E_F = -58 meV$ (solid curve). The inset displays the nesting effect of the Fermi surface at the wave vector X $= \frac{\pi}{a}\left(1,1,0\right)$. $\Delta \sim \left(1,0,0\right), \Sigma \sim \left(1,1,0\right)$. The broken lines display the dominant matrix element $\Pi_{KK}\left(\vec{q}\right)$ according to Eq. (11) related to the *Cu 3d* orbital for $E_F = -58 meV$ (van Hove case). *a* is the lattice constant.

In the following we investigate the question whether Fermi-surface nesting could provide an alternative explanation for the phonon anomalies. According to [68] a very strong renormalization of $O_b^X$ caused by Fermi-surface nesting effects leading to gigantic *Kohn anomalies* has been predicted and so we have recalculated the situation within the 31BM. It is well known that strong nesting at some wave vector $\vec{q}$ or large matrix elements at $\vec{q}$ or both may lead via an increased polarizability $\Pi(\vec{q})$ to a softening for certain phonon modes at this wave vector. A possible measure that could indicate the tendency toward such Fermi-surface driven anomalies is the scalar polarizability

$$\Pi_o(\vec{q}) = -\frac{2}{N}\sum_{\substack{n,n'\\ \vec{k}}}\frac{f_{n'}(\vec{k}+\vec{q})-f_n(\vec{k})}{E_{n'}(\vec{k}+\vec{q})-E_n(\vec{k})} \qquad . \qquad (23)$$

The structure in $\Pi_o$ depends mostly on parallel sheets of the Fermi-surface, with a stronger nesting leading to sharper structure. The inset in Fig. 10 displays the nesting effect for the wave vector $\vec{q} \cong \vec{X} = \frac{\pi}{a}(1,1,0)$ within the 31BM for the Fermi energy at $E_F = 0$ and $E_F = -58 meV$, respectively. The latter case corresponds in a rigid band model to optimally doped $La\text{-}Cu\text{-}O$ where $E_F$ lies at the van Hove singularity of the density of states. Moreover, in Fig. 10 the intraband contribution $(n = n')$ of $\Pi_o$ from Eq. (23) is shown for $E_F = 0$ (dotted curve) and the van Hove case (full curve). For comparison also the dominant matrix element $\Pi_{\kappa\kappa'}(\vec{q})$ of the $Cu\ 3d$ orbital of the full polarizability matrix given in Eq. (11) is plotted (broken curve) for $E_F = -58 meV$. From an inspection of Fig. 10 a nesting effect in $\Pi_o(\vec{q})$ can only be discerned around the $X$ point. For $E_F = 0$ there are additionally weak features at different wave vectors. In the van Hove case the peak in the polarizability is enhanced and slightly shifted away from $X$ towards $\Gamma$. On the contrary, in the middle of the $\Delta$ direction where the phonon anomalies are strongest $\Pi_o(\vec{q})$ approaches a minimum. Thus, there is no explanation of the strong phonon renormalization along the $\Delta$ direction by nesting of the FS. The softening of $O_B^X$ is primarily also not related to FS nesting, because the nesting effect seen in $\Pi_o(\vec{q})$ is strongly suppressed by the components of the eigenvectors $C_{\kappa n}(\vec{k})$ contributing to the full matrix $\Pi_{\kappa\kappa'}(\vec{q})$, see the broken curve in Fig. 10 representing the dominant matrix-element $\Pi(Cu\ 3d, Cu\ 3d)$ of $\Pi_{\kappa\kappa'}(\vec{q})$. The remaining matrix-elements $\Pi_{\kappa\kappa'}(\vec{q})$ are even an order of magnitude smaller [69]. We note that in $YBa_2Cu_3O_7$ there is also no indication that nesting may be the source for the generic phonon anomalies found also in this material in our calculations. However, changes of the FS from rounded squares oriented in the (1,1,0) directions as in $La\text{-}Cu\text{-}O$ (Fig. (10)) to rounded squares in the (1,0,0) directions in $Y\text{-}Ba\text{-}Cu\text{-}O$ lead e. g. to small wave-vector nesting between different pieces of the FS and corresponding peaks in $\Pi_0(\vec{q})$ [69] which may enhance scattering in the *antinodal* region while $\Delta_1/2$ enhances scattering in the *nodal* region.

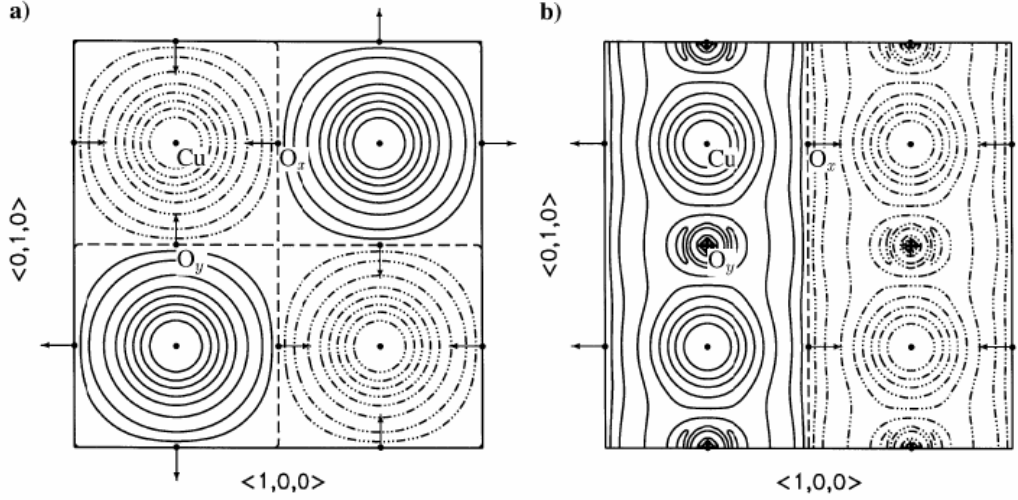

**Fig. 11** Contour plot of the nonlocal part of the phonon-induced charge density redistribution $\delta\rho$, according to Eq. (24) for (a) the $O_B^X$ mode and (b) the $\Delta_1/2$ mode. The displaced ions are indicated by the dots with the arrows. $\delta\rho > 0$ (full lines) means that electrons are accumulated in the corresponding region of space. Broken lines $(-\cdots)$ represent the regions where the electrons are pushed away. At the line $(---)$ $\delta\rho = 0$. $\delta\rho$ is given in units of $\left(10^{-4} e/a_B^3\right)$ and the different lines correspond to the values: $0, \pm 0.05, \pm 0.15, \pm 0.5, \pm 1.25, \pm 2.5, \pm 5, \pm 10, \pm 25$.

Fig. 11 displays our calculated results of the nonlocal part of the phonon-induced charge density redistribution

$$\delta\rho_{nl}(\vec{r},\vec{q}\sigma) = \sum_{\vec{a}\kappa} \rho_\kappa\left(\vec{r} - \vec{R}_\kappa^{\vec{a}}\right)\delta\zeta_\kappa^{\vec{a}}(\vec{q}\sigma) \tag{24}$$

for $O_B^X$ (a) and $\Delta_1/2$ (b) for the optimally doped metallic phase with $\delta\zeta_\kappa^{\vec{a}}$ from Eq. (12). For the calculation the same model is used as for the phonon dispersion shown in Fig. 8. The full lines indicate the regions in space where the electronic charge density increases and the broken lines stand for a decrease. During the $\Delta_1/2$ mode the moving $O_x$ ions generate via nonlocal

EPI changes of the potential at the silent $Cu$ and $O_y$ ions resulting in corresponding CF's in form of a charge transfer within and between the $CuO_y$ chains as shown in Fig. 11 (b). Note that there are no changes of the transfer integral between $d$ and $p$ orbitals for the silent $Cu$ and $O_y$ ions, nevertheless there is a charge transfer. In case of the $O_B^X$ mode the moving $O_{xy}$ ions induce CF's at the silent $Cu$ and we obtain an electronic charge transfer from that $Cu$ ion where the $CuO$ bonds are compressed to the $Cu$ where the $CuO$ bonds are stretched, see Fig. 11 (a).

According to these calculations the nonlocal EPI leads to dynamic charge ordering by CF's in the form of localized stripes of alternating sign in the $CuO$ plane which are strongly interacting with the lattice vibrations and with each other. In case of the $\Delta_1/2$ mode the charge stripes point along the $x$ or $y$ axis, respectively, and for $O_B^X$ along the diagonals in the $CuO$ plane. The charge patterns appear instantaneously because the adiabatic approximation is sufficient for these modes, i. e. the excited CF's are at a higher energy scale than that of the phonons, see, however, Part 3.4 for a non-adiabatic scenario for modes propagating along the $c$-axis.

Quite generally, the results visualized in Fig. 11 demonstrate the characteristic property of the charge response in the HTSC's to develop excitations in form of localized CF's, i.e. dynamic charge inhomogeneities, under perturbations, where the character of the latter changes qualitatively upon doping from more localized with a higher energy ($Cu\ 3d$ type) to a situation where the weight of the delocalized CF's with a lower energy ($O\ 2p$-, $Cu\ 4s$ type) increases. Such a specific doping dependence of the localization properties of the electronic structure, leading to this specific type of screening, of course has influence on the movement of the charge carriers themselves, not only in the $CuO$ plane but also along the c-axis. Thus the carriers will not only be expected to couple to a continuum of SF's, as is commonly assumed, but also nonlocally to the phonons and to a continuous spectrum of CF's, where the SF's and the more localized CF's should contribute to the broad band observed in optical spectra of the HTSC's in the $CuO$ plane extending beyond the cut-off frequency of the phonons and the delocalized low-energy CF's to the Drude component, as already remarked. As far as isotope effects are concerned the electronic selfenergy can be expected to depend on the oxygen isotopes via the nonlocal EPI-effects of CF type accompanied by SF's, e.g. mediated by the anomalous OBSM, and also on the non-adiabatic phonons around the c-axis in a generic fashion. So, our calculations for the OBSM illustrate how phonons, CF's and SF's in total are at work in the normal and superconducting state of the HTSC's.

It is quite interesting to note that electron correlation effects generated by strong repulsive Coulomb interactions which result in an antiferromagnetic correlation of sufficient correlation length may favour the CF's in the modes just discussed and enhance screening against the tendency of a suppression of the CF's by the on-site repulsion at the $Cu$, because if the spins in neighbouring $Cu$ orbitals were to be parallel, they would not both be able to transfer to the neighbouring oxygen or between the $Cu$ ions themselves. On the other hand, if they are antiparallel they can. Thus, the phonon induced CF's depend implicitly on the spin-degrees of freedom (at least in a picture where charge and spin are not separated) which enhance the former and the EPI in an antiferromagnetic neighbourhood as compared to a paramagnetic or a ferromagnetic one. Vice versa the nonlocal EPI is expected to generate antiferromagnetic SF's via the phonon-induced transferred charge between the (spin-polarized) $Cu$ ions, see Fig. 11. For this process the renormalization of hopping of the charge carriers and possibly a certain change of the exchange coupling by EPI should be considered in an advanced many body treatment. In this way the phonon dynamics and the CF's can renormalize the spin dynamics and vice versa. Moreover, the induced CF's and SF's should match in frequency to minimize spin frustration. Summarizing, spin-degrees of freedom and their phonon induced fluctuations may enter the CF-scenario driven by the high-frequency OBSM and the discussion of the physics around these modes provides an example to illustrate, how coupled lattice-, charge- and spin-degrees of freedom could act synergetically in the doped cuprates. This interplay ultimately results in high-temperature superconductivity with highest $T_c$ in the optimally doped metallic phase of these complex materials where the doping dependent changes in localization are an important aspect.

To be more specific as far as the phonon contribution via nonlocal EPI is concerned the most favourable situation for binding Cooper pairs is apparently when two electrons (holes) interact with a time-retardation $1/(4\nu)$ ($\nu$: phonon frequency). Such a situation can be visualized for example by means of the $\Delta_1/2$ anomaly shown in Fig. 11b. Here the "first" electron moving along a $Cu\,O_y$ chain pushes via nonlocal EPI the neighbouring $O_x$ ions away and generates thereby characteristic CF's $\delta\zeta_\kappa^{\vec{a}}(\vec{q},\sigma)$ and corresponding changes in potential $\delta V_\kappa^{\vec{a}}(\vec{q}\sigma)$ for its time-retarded partner in the pair. Note, that in such a mechanism we have both, nonlocality in time as usual but also nonlocality in space which is not considered in the conventional pairing mechanism mediated by phonons. The nonlocality in space offers the possibility to anisotropic pairing states by allowing the Cooper pair to sample mainly the attractive regions of the interaction. For example our calculations of the $\delta V_\kappa$ in the $\Delta_1/2$ mode for the optimally doped phase have shown that in case the $O_x$ ions are pushed away from the $Cu\ O_y$ chain (Fig. 2) the space region around $O_y$ is attractive and that around $Cu$ repulsive for electrons which is favourable e.g. for *d-wave pairing*. Analogous considerations about pairing due to nonlocal EPI also hold in case of the $O_B^x$ mode for the Cu-chains in (1,1,0) direction, see Fig. 11a. It is very likely that the $\Delta_1 / 2$ mode is more important if the mobile charge carriers (holes) are doped into the oxygen orbitals as in *p-type* cuprates. In this context also the *phase space* is important where strong nonlocal EPI leads to the phonon anomalies and the favourable situation for pairing. It has been shown [69] that the latter is of considerable size, see below. Comparing the underdoped phase with the optimally doped phase we can expect a different nature of superconductivity from our modelling, because in case of the optimally doped phase the "*correlation gap*" vanishes and besides the *O 2p* states the *Cu*-related states additionally become metallic in the normal state and this makes metallic CF's possible also at the *Cu* sublattice. In other words, the loss in density of states at $E_F$ related to the *Cu* states in the underdoped material is restored upon doping and thus, these states additionally become available for pairing via charge degrees of freedom and not only via virtually hopping as in the underdoped case.

Finally, the OBSM synergetical for pairing, may also influence the antiferromagnetic spin resonant excitations that develop below $T_c$ in the HTSC's for which electron-electron interactions most likely play the central role, see Ref. [70] for a recent overview. Such a conjecture might be expected from inspection of Fig. 11 because the phonon-induced charge transfer between the *Cu* ions is accompanied by corresponding spin-fluctuations. Moreover, on account of the superconducting gap below $T_c$ (which opens along (1,0,0) and closes along (1,1,0) in case of *d-wave symmetry*) EPI-effects via low-lying optical phonons, are partially blocked and do not interfere with the constructive effects provided by the OBSM, so essentially the latter modes with their high frequencies and considerable weight in phase space are active favouring spin excitations. Thus, the OBSM seem to be synergetical for both, pairing and an enhancement and renormalization of the spin excitations in the superconducting phase. The latter represent a feedback effect related with the onset of superconductivity. As speculated in [70] and recently found experimentally in [71] by an analysis of the optical self-energy these resonant magnetic excitations seem to disappear completely in the overdoped regime at a doping level where superconductivity is still large. Thus, if true the magnetic resonance peak cannot be the basic reason for high-$T_c$ superconductivity and the remaining source left for pairing are CF's, phonons and incoherent SF's, where the latter vanish in parallel with the weakening of the many-body effects at higher doping, as seen in the experiments.

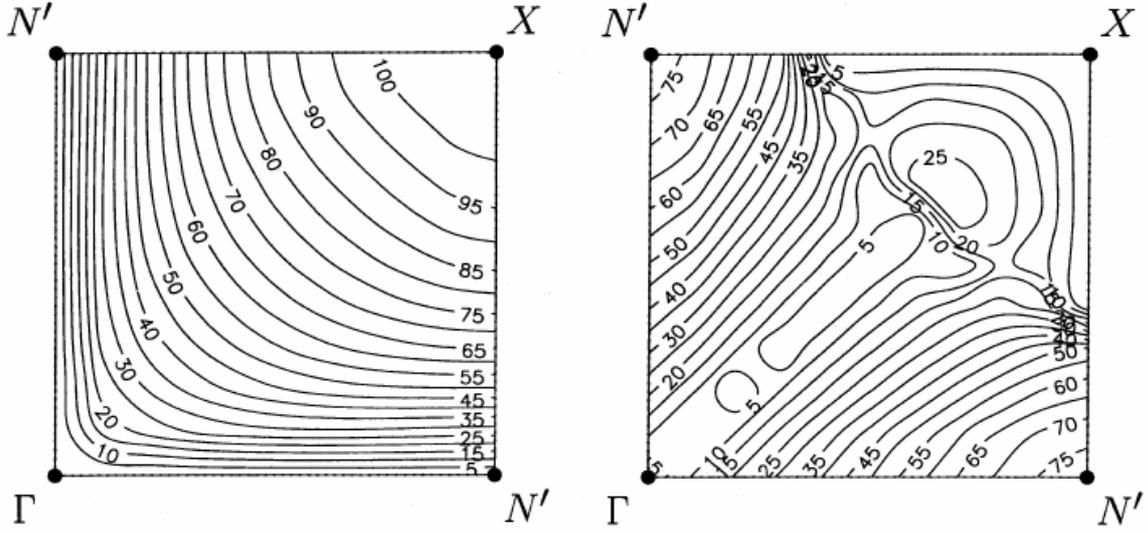

**Fig. 12** Contour plot of the phonon-induced self-consistent changes of the crystal potential (in meV) of *La-Cu-O* from Eqs. (15 – 17) for the dominant *Cu 3d* orbital in the $q_Z = 0$ plane. In the left part the result for the branch with the $O_B^X$ anomaly at the *X* point is given and in the right part the result for the branch with the $\Delta_1 / 2$ anomaly at $N' = \frac{\pi}{a}\left(1, 0, 0\right)$.

Next we explore the size of phase space where the strong nonlocal electron-phonon coupling appears. We will see, that the strong coupling effects leading to the phonon anomalies in *La-Cu-O* are not restricted to a small portion of the BZ but extend to a considerable part. As a measure of the strength of the EPI in a certain phonon mode we apply the phonon-induced changes of the self-consistent potential $\delta V_\kappa\left(\vec{q}\sigma\right)$ an electron feels, defined by Eqs. (15) – (17). The result for $\delta V_\kappa\left(\vec{q}\sigma\right)$, [69], is shown in Fig. 12 in the $q_z = 0$ plane of the BZ for the *Cu 3d* orbital. In the left part of Fig. 12 the data for $\delta V_\kappa$ of the phonon branch with the highest frequency in the spectrum are given with the $O_B^X$ anomaly at the $X = \pi / a\left(1, 1, 0\right)$ point. The right part of Fig. 12 displays the result for $\delta V_\kappa$ of the second highest branch with the $\Delta_1 / 2$ anomaly at $N' = \frac{\pi}{a}\left(1, 0, 0\right)$. Our calculations further have shown that the anomalous modes practically have no dispersion in the direction perpendicular to the $q_z = 0$ plane. This means that the strong coupling also extends to the $q_z$ direction of the BZ. Altogether, the calculations demonstrate that in a considerable part of the BZ around the $\Delta_1 / 2$ anomaly and $O_B^X$ , respectively, strong nonlocal phonon-induced changes in the potential can be found.

In [69] we also have investigated the problem of an anomalous phonon softening of the OBSM for $Y\,Ba_2Cu_3O_7$ . From our results we extract, likewise as in *La-Cu-O*, a strong softening of these modes which now, however, have doubled in number because of the two *CuO* layers, which is an advantage of the two-layer material as compared with the single-layer case concerning the pairing contribution via nonlocal EPI. The physical reason for the strong electron phonon coupling is the same, namely nonlocal EPI of ionic charge-fluctuation type in the *CuO* planes. This supports the generic nature of the phonon anomalies in the HTSC's. A certain drawback of our calculations for $Y\,Ba_2Cu_3O_7$ is the fact that the RIM, which serves as the unbiased reference system for the investigation of the non-rigid screening effects by the CF's and DF's, predicts for the OBSM frequencies which are somewhat too large as compared to the experimental results. Thus, while the phonon renormalization of the OBSM by CF's can be studied quantitatively by comparing with the corresponding results for the RIM, other interesting aspects of the phonon dispersion of $Y\,Ba_2Cu_3O_7$ currently under active experimental research like possible *anticrossing effects* with lower lying branches of the same symmetry but different polarization are (at least partly) not directly visible in the calculations. Such an anticrossing of the OBSM in $Y\,Ba_2Cu_3O_7$ with other branches of the same symmetry is, very likely according to recent experimental results [72]. In this work anticrossing of the OBSM in (1,0,0) direction with a *c*-axis polarized branch which starts from an $A_g$ mode at the $\Gamma$ point at *62 meV* is reported. Further complication arises in the (0,1,0) direction from the presence of another branch of the same symmetry in the energy region of interest (longitudinal *Cu-O* chain bond-stretching mode at $\Gamma$ ). While the anticrossing with the oxygen chain mode has been predicted in [69] the anticrossing with the *c*-axis polarized branch is not visible in the calculations.

However, shifting the calculated frequencies of the OBSM to the energy range of the experimental results an anticrossing of the OBSM with a *c*-axis polarized branch starting in our calculations at $\Gamma$ at about *64 meV* in form of a an $A_g$ apex-oxygen mode becomes evident. This fact supports the anticrossing arguments given in Ref. [72]. The existence of the anticrossing effects in $Y\,Ba_2Cu_3O_7$ over a substantial part of the BZ is important for an understanding of the complex lineshapes observed for the OBSM. Moreover, as discussed in [72] the anticrossing effect could explain the "*branch splitting*" of the OBSM branch reported in [73] in a natural way, where this "splitting" has been interpreted speculatively as a signature of charge stripes in the *CuO* planes.

From our studies of the OBSM we have found that doping dependent changes of the localization properties of the electronic states correlate critically with the strength of the phonon anomalies and simultaneously with the superconducting properties. So, in a multilayered HTSC's where different doping levels are realized in the single layers as reported in recent nuclear magnetic resonance measurements of the *Hg*-series [74] the localization properties in the planes differ as well. In [74] it is found that the outer layers are predominantly overdoped whereas the inner layers are underdoped and this may be the cause why $T_c$ does not rise with the numbers of layers but reaches a maximum at *n=3* and then declines. Recently an explanation for such a behaviour has been proposed within the framework of Ginzburg-Landau theory by studying competing order of superconductivity with another order parameter related to the charge imbalance between the layers assuming interlayer pair tunnelling as an additional mechanism by which pairing could be enhanced [75]. From our investigations an optimal $T_c$ can be expected if the critical localization properties of the states found in the calculations of the OBSM for the optimally doped state (Figs. 8, 9) are realized in *all* the planes of the multilayer system and charge imbalance could be avoided. In this context, in particular the *Cu 4s* occupation controlling intralayer hopping beyond nearest neighbours as well as interlayer hopping is an important material dependent quantity to optimize superconductivity. Deviations from these sensitive localization conditions will tend to reduce $T_c$ and adding planes will not automatically improve superconductivity, because as already mentioned in the underdoped regime the localization of the *Cu 3d* states and the corresponding "*correlation gap*" leads to a loss of density of states at $E_F$ and these states are no longer available for pairing via charge degrees of freedom. On the other hand, in the overdoped regime the strong nonlocal EPI is weakened and the SF's die out because of the increased delocalization; both leads to a weaking of superconductivity. Thus, in order to achieve a maximal $T_c$ it is necessary to avoid charge imbalance in the doping process, for example by trying to modify the ionic layers in the HTSC's.

We now draw attention to other theoretical attempts relying on simplified models with reduced model Hamiltonians and restricted orbital degrees of freedom as well as estimated parameters that do not incorporate the long-ranged Coulomb interaction to understand the phonon anomalies in the HTSC's and their possible role for high-temperature superconductivity. In [76] a two-band Peierls-Hubbard model in one dimension is used to study zone-boundary *LO* phonon modes and it was found that the latter couple very strongly to holes resulting in charge transfer and changes of the local spin correlation by the deformation. These effects are proposed to explain the softening of the $(\pi, 0) - LO$ phonon by doping. Strong coupling of the Cu- and O-half-breathing mode to doped holes is reported in [77] in the framework of an effective two-band t-J model. A phonon mechanism of high-temperature superconductivity due to phonon-induced charge transfer by the OBSM is proposed in [78]. In [10, 79] it is found that the oxygen half-breathing mode strongly contributes to *d-wave pairing*. The calculations in these papers have been performed in a *t-J* model assuming off-diagonal electron lattice coupling as the dominant mechanism. On the other hand, in the studies [80] within the *t-J* model anomalous softening of the $\Delta_1 / 2$ mode upon doping can be found assuming that off-diagonal electron-phonon coupling is small and the diagonal electron-phonon coupling dominates in the cuprates. The calculations [81] within a three-band Peierls-Hubbard model suggest that the softening of the oxygen half-breathing mode in both *p* and *n-type* cuprates can be attributed to the strong interactions of lattice distortions with local metallic CF's. In addition it is shown that the spin-phonon coupling is much stronger in the *p-type* cuprates than in the *n-type* cuprates.The investigations in [82] using a two-dimensional one-band Hubbard model for a study of the electron-phonon vertex correction show that in the weak- and intermediate coupling regimes the on-site Coulomb interaction suppresses the ionic electron-phonon coupling. However, in the strong-coupling regime at a sufficiently large on-site interaction the EPI is enhanced at small phonon momentum transfer. Such finding already has been predicted in [83] by an approximate treatment of the Hubbard model. In addition, an EPI which is peaked at small phonon momentum transfer contributes to an attractive interaction for *d-wave pairing* [83, 84]. An early study how strong Coulomb correlations affect the phonon-mediated superconductivity is presented in [85] and an interesting review on high-temperature superconductivity focusing in particular on the phonon aspect is given in [86]. The increase of the EPI as a function of the on-site interaction is attributed to the renormalization of the low-energy single-particle and charge excitations [82]. A quasi-particle renormalization [29] resulting in flat bands at $E_F$ enhances the density of states at the Fermi level, this, in tendency decreases $\Pi^{-1}$ in Eq. (9) and leads for a fixed $\tilde{V}$ to a reinforcement of the density response function, $C^{-1}$, in Eq. (10). The flat bands are unfavourable for energy relaxation via the kinetic part of the total energy. According to Eqs. (9), (10) the effect of $\tilde{V}$ on the charge response, the EPI and the phonon frequencies is relatively enhanced as compared to that of $\Pi^{-1}$; in particular if one is already in the strong coupling regime. In detail, of course, things depend on the way $\Pi^{-1}$ is changed as a function of $\tilde{V}$. Qualitatively, in the regime of strong correlations the kinetic contribution of the single particles to $C\left(C^{-1}\right)$ is not as important as that of $\tilde{V}$ because strong electron-repulsion inhibits the stabilizing role for the kinetic energy. On the other hand for delocalized states with a smaller on-site interaction energy relaxation via the kinetic energy becomes significant and plays an important role also for the additional softening of the OBSM in the optimally doped phase as discussed earlier in this section. Altogether, many of these recent model considerations point to the result that the anomalous softening of the OBSM in the doped metallic phase is driven by strong nonlocal EPI of localized, ionic CF-type, already predicted by our calculations [34, 45, 58] in quantitative agreement with the inelastic neutron scattering measurements.

Quite recently we also have found in our calculations a strong renormalization of the high-frequency OBSM in the $K$-doped non-cuprate perovskite high-termperature superconductor $Ba\text{-}Bi\text{-}O$ $\left(T_C \sim 30K\right)$ generated by strong nonlocal EPI of CF-type similarly as in the cuprate-based HTSC's. $K$-doping in $Ba_{1-x}K_xBiO_3$ creates holes on the $O2p$ orbitals and on crossing the insulator-metal transition these holes become itinerant and contribute to the metallic behaviour. According to our calculations the anomalous softening of the modes in the doped metallic and superconducting phase is essentially a nonlocal coupling effect of the (metallic) $Bi\ 6s$ charge-fluctuations to the displaced oxygen ions where the relatively small on-site Coulomb interaction for the $Bi\ 6s$ electrons and a large $Bi\ 6s$ polarizability enhances the CF's at the $Bi$ site and finally leads to the strong renormalization. Doped $Ba\text{-}Bi\text{-}O$ is different from the HTSC's cuprates since no antiferromagnetic ordering exists for the insulating parent compound. While HTSC's are nearby an insulating ionic-antiferromagnetic groundstate, superconducting $Ba\text{-}Bi\text{-}O$ is in the neighbourhood of an insulating ionic-charge-density-wave groundstate in which oxygen octahedra around the $Bi$ ions show alternating breathing-in and breathing-out distortions. Common to both materials is the strong component of ionic binding favouring localized CF's and the long-ranged Coulomb interactions. Thus, while spin-degrees of freedom seem to play no decisive role for superconductivity in $Ba\text{-}Bi\text{-}O$ the effects of nonlocal ionic coupling of lattice- and charge-degrees of freedom in terms of CF and DF are still present and provide coupling channels for the pairing mechanism also in this material. The result of our calculations of an *anomalous* and *anisotropic* softening of the OBSM in $Ba\text{-}Bi\text{-}O$ is displayed in Fig. 13 for the $\Delta \sim \left(1,0,0\right), \Sigma \sim \left(1,1,0\right)$ and $\Lambda \sim \left(1,1,1\right)$ direction [87].

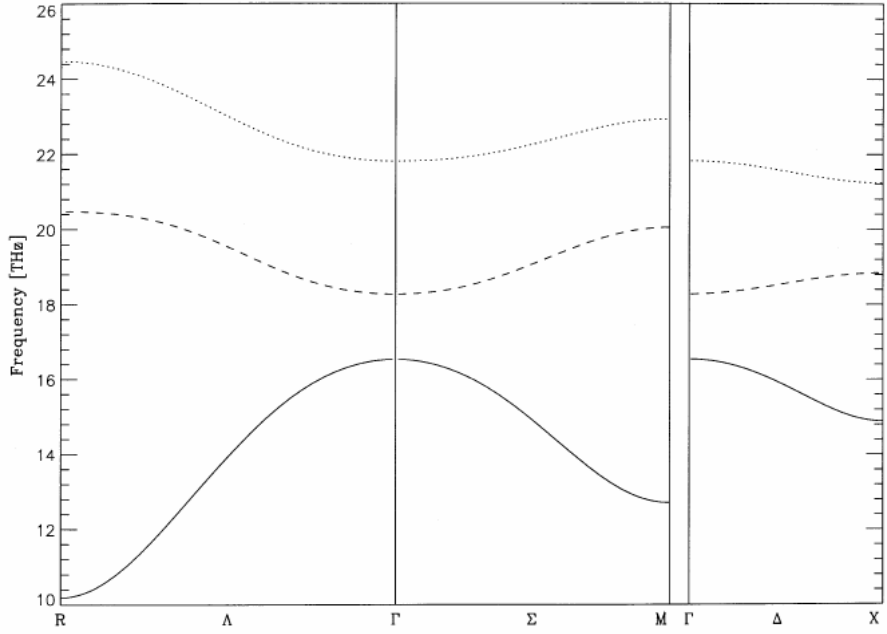

**Fig. 13** Calculated anomalous branches of the high-frequency oxygen bond stretching modes of metallic *Ba-Bi-O* along the $\Lambda \sim (1,1,1)$, $\Sigma \sim (1,1,0)$ and $\Delta \sim (1,0,0)$ direction with the oxygen volume breathing mode at R, the planar oxygen breathing mode at M and the linear oxygen breathing mode at X. The linetype •••• denotes the result for the RIM, − − − includes additional to the RIM DF's and the full lines represent the calculation including DF's as well as CF's.

A particular strong renormalization is predicted by our calculations for the oxygen volume-breathing mode at *R*. The mode softening from $\Gamma$ to *R* amounts to about *6.4 THz* that is about *40 %*. This is the strongest softening so far in any superconductor. The planar oxygen breathing mode at *M* is renormalized by about *3.8 THz* and the linear breathing mode at *X* by about *1.7 THz*. Our predictions shown in Fig. 13 can be compared with results from inelastic neutron scattering experiments [88] where strong anisotropic phonon anomalies for the OBSM are obtained for doped *Ba-Bi-O*. Likewise as in our calculations a giant phonon anomaly at the *R* point by about *40 %* is observed with a frequency near *10.5 THz* which is in excellent agreement with our calculation (*10.2 THz*).

As a final comment concerning the broad range of applicability of our microscopic model we would like to point out that the formalism recently also has been used successfully for the calculation of the phonon dispersion, dielectric properties and in particular for a comparative study of the microscopic origin of the structural instabilities (ferroelectric and antiferrodistortive) in $BaTiO_3$ and $SrTiO_3$, respectively [89].

### 3.4 Non-adiabatic *c*-axis phonons and phonon-plasmon mixing

The nature of the *c*-axis charge response of the HTSC's is still not well understood. Bandstructure calculations for the cuprates within DFT-LDA always predict an appreciable *c*-axis dispersion and thus an anisotropic three-dimensional metallic state [18] which underestimates the anisotropy along the *c*-axis. In contrast to this Fermi-liquid-based approach, the Luttinger-liquid model proposes that the charge carriers are strictly confined to the *CuO* planes [90]. However, a strictly two-dimensional model a priori neglects the three-dimensionality of the real materials resulting from direct Coulomb interaction and dielectric coupling between the layers. These long-ranged interactions have been shown by our calculations in the preceding Subsections to be important for a realistic description of the electronic density response, EPI and lattice dynamics. Moreover, quite recently in the *Bi-2212* cuprate a bilayer splitting has been reported in photoemission experiments in good agreement with bandstructure calculations [91 – 93]. This implies evidence for the lack of any electronic confinement to single planes due to strong correlations in contrast to the confinement assumption in [90]. As mentioned in Subsection 3.2 a phonon-induced bilayer charge transfer has been generated in *Bi-2212* by the $A_{2u}$ modes consistent with an electronic coupling between the *CuO* planes. Moreover, as our preliminary calculations within a non-adiabatic scenario for *Y-Ba-Cu-O* indicate [94], a weak intra-bilayer coupling generates a partial reduction of the *LO-TO* splitting and modifies the optical *c*-axis response. According to these calculations the intra-bilayer coupling produces a low-lying *interband-plasmon* due to interband transitions mixing with the *c*-axis phonons. These effects show up, besides the low-lying interband transitions, in the dynamical *c*-axis charge response. So, we have besides the *intraband-plasmon* also found in a single layer material (see below) and the doubling of the strongly coupling OBSM the *interband-plasmon* coupling with the *c*-axis phonons as an additional electronic resonance and a coupling channel for pairing in the bilayer material. Note in this context, that the *intraband-plasmon* generated by interlayer coupling via intraband transitions will interfere with the interband-plasmon and that increasing interlayer coupling, i. e. reducing the anisotropy, will tend to screen the interband transitions and *LO-TO* splitting ultimately will vanish, see also the discussion below.

A model resting on the Fermi-liquid picture, proposed in an effort to classify the HTSC's as regards their *c*-axis transport explains the experimental situation quite well [95]. Here the out-of-plane charge transport comes from an interplay of a coherent component (direct interlayer coupling) and incoherent contributions. The relative strength of both contributions varies for the HTSC's depending on their degree of anisotropy.

Following our discussion given in reference [32] the effect of the anisotropy on the phonon dispersion of $La_2CuO_4$ along the $c$-axis and in a small region around this axis is studied in terms of the magnitude of the coherent component of the $c$-axis charge response. We have investigated phonon-plasmon mixing along the $\Lambda \sim (0,0,1)$ direction for the relevant modes starting from the non-adiabatic soft-plasmon case of a two-dimensional quasiparticle bandstructure and going up to the adiabatic limit of a less anisotropic metal as predicted typically by bandstructure calculations within DFT-LDA. For a discussion of phonon-plasmon mixing in the cuprates in the context of many-body polaronic effects in the phonon spectrum, we refer to [96].

As far as optimally doped and underdoped $La$-$Cu$-$O$ is concerned, the optical $c$-axis spectra display features typical of an ionic insulator [97 – 99] and not of a metal. They are dominated by optical phonons and are almost unchanged from those of the insulator upon doping. On the other hand, the $c$-axis neutron data seem to contradict the optical $c$-axis results. While the phonon dispersion of undoped $La_2CuO_4$, see. e. g. Fig. 5, also shows characteristic $A_{2u}$ mode discontinuities typical of ionic insulators, in the doped metallic samples these splittings disappear and a phonon dispersion results along the $\Lambda$ direction, which is typical for a three-dimensional anisotropic metal as calculated in the adiabatic approximation, see Fig. 4. Connected with the disappearance of the $A_{2u}$ discontinuities in the metallic phase is the appearance of the characteristic phonon branch of $\Lambda_1$ symmetry with the very steep dispersion already mentioned in Subsection 3.1. This signature is predicted also by our calculations in the adiabatic approximation for the metallic phase in Fig. 4.

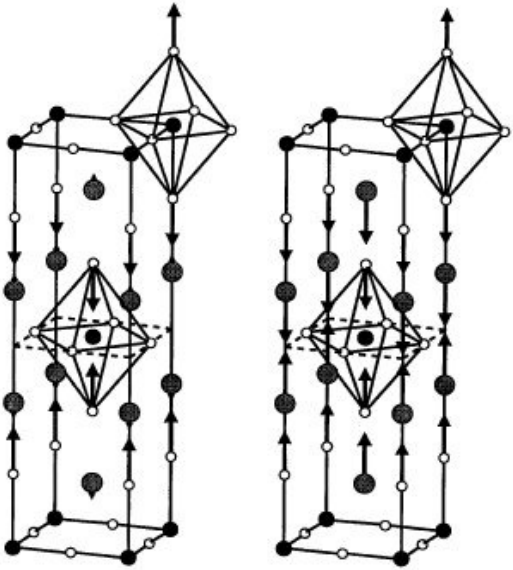

**Fig. 14** Displacement pattern of the symmetrical apex-oxygen breathing mode, $O_z^Z$ (left) and the symmetrical La-breathing mode, $La_z^Z$ (right), at the $Z$ point of the BZ.

An explanation of the apparent inconsistency between the current interpretation of the neutron scattering results for the $\Lambda_1$ branches in doped metallic samples, as in Fig. 4, and the infrared data being typical for an ionic insulator will be presented below within a non-adiabatic phonon-plasmon scenario. Moreover, this scenario also allows to understand the large softening (about *6 THz*) and the massive line-broadening (about *4 THz*) of the symmetrical apex-oxygen-breathing mode, $O_z^Z$, at the *Z* point (Fig. 14) experimentally observed [3], when going from insulating to optimally doped metallic *La-Cu-O*. Altogether, these results support a phonon-plasmon scenario around the *c*-axis in the HTSC's. In the overdoped material with an increasing interlayer coupling the *c*-axis plasmon frequency increases in parallel and the charge response will become more and more adiabatic, as can be seen from the metallic resistivity curves and the optical response in such probes [63, 97].

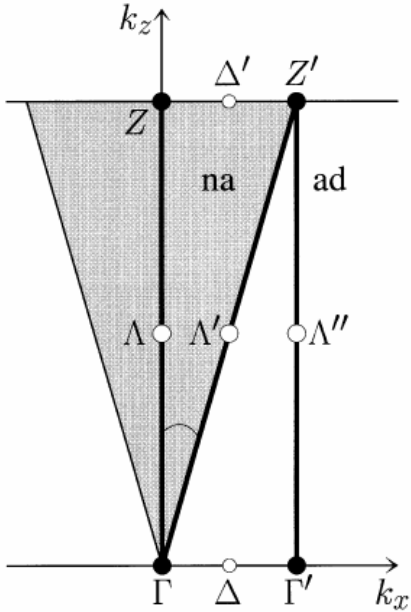

**Fig. 15** Schematic representation of the non-adiabatic region in the $\left(k_x k_z\right)$-plane with the directions

$$\Delta' = \left(\varepsilon \frac{2\pi}{a}, 0, \frac{2\pi}{c}\right), \Lambda' = \zeta\left(\varepsilon \frac{2\pi}{a}, 0, \frac{2\pi}{c}\right) \text{ and }$$

$$\Lambda'' = \left(\varepsilon \frac{2\pi}{a}, 0, \zeta \frac{2\pi}{c}\right); \zeta \in [0,1].$$ na: non-adiabatic region; ad: adiabatic region

The loss of interlayer coupling with increasing anisotropy leads to a slow electron dynamics and has the consequence that screening of the bare long-ranged EPI becomes ineffective in a certain region around the *c*-axis to be estimated below (non-adiabatic region in the following, see. Fig. 15) and forward-scattering dominates. This results in a strong enhancement of the EPI for the coupled phonon-plasmon modes in this region. In this context *Migdal's theorem* does not apply and one has to take into account these non-adiabatic effects in a description of superconductivity.

As remarked in [32] there is also a relation between strong correlations in the *CuO* plane and the size of the non-adiabatic region pointing to an interplay of non-adiabatic effects and strong electronic correlations. A correlation induced reduction of the bandwidth and the Fermi velocity of the single-particle excitations in the $(k_x k_y)$-plane compared with DFT-LDA-like bandstructures will increase the size of the non-adiabatic region around the $c$-axis. This is because the free-plasmon frequency is decreased in such a case following equations (10, 11, 18). Quite recently such an effect also has been studied numerically [42]. In this way the characteristic properties of the QP in the *CuO* plane related for example to strong correlations take influence on the dynamical charge response perpendicular to the plane nearby the $c$-axis. In order to study the anisotropy effect on the phonon-plasmon dynamics we have assumed in [32] an eleven-band model (11 BM) for the description of the single-particle excitations in the *CuO* planes and extended this model by a suitable interlayer coupling in parameterized form. Such a model then has been used in the expression for the dynamical version of the electronic polarizability $\Pi$ according to Eq. (11). In addition anisotropic DF's on the ions along the *z*-direction are taken into account and the dipole polarizability has been calculated by the Sternheimer method within DFT-LDA-SIC.

By increasing the interlayer coupling $K$ (measuring the strength of the coherent charge response) over a sequence of five models, M1 – M5, we study the effect of the anisotropy and the position of the $c$-axis plasmon, separately, on the coupled phonon-plasmon modes of $\Lambda_1$ symmetry by solving Eq. (5) for the dynamical matrix selfconsistently. The $z$-polarized $\Lambda_1$ branches are the only phonons that couple strongly with the CF [32, 46]. We shall demonstrate that whether one observes metallic screening effects of the $\Lambda_1$ modes depends on the frequency of the $c$-axis plasmon in comparison to the phonon frequencies. Thus, the dispersion may look like that of an insulator or that of a metal, respectively.

Model M1 in Fig. 16 corresponds to $K$=0 and represents the limit of a strictly two-dimensional quasiparticle (QP) band-structure. The $c$-axis plasmon is soft ($\omega$=0) because in this extremely anisotropic case there is no coherent contribution to the charge transport along the $c$-axis. Intraband quasiparticle screening along the $c$-axis is absent in this limit and the physics is governed by higher energy interband charge transfer processes like in an insulator leading to large long-ranged EPI. At the same time an *acoustic plasmon* mode develops when proceeding from the $\Lambda$ direction to the $(q_x q_y)$-plane in the BZ [30]. In the other models M2 – M5 with decreasing anisotropy, i. e. increasing interlayer coupling, we have chosen $K$ in such a way that the free (now *massive*) plasmon at $\Gamma$ in the extended 11 BM is about *1THz* (M2), *5THz* (M3), *10THz* (M4) and *18THz* (M5); see Table 4. From this table we can also deduce that the interband screening effect of the dielectric coupling between the layers in terms of anisotropic DF's considerably reduces the frequency of the free plasmon in the extended 11 BM. This again demonstrates the importance of the three-dimensionality for the electronic properties in the HTSC's.

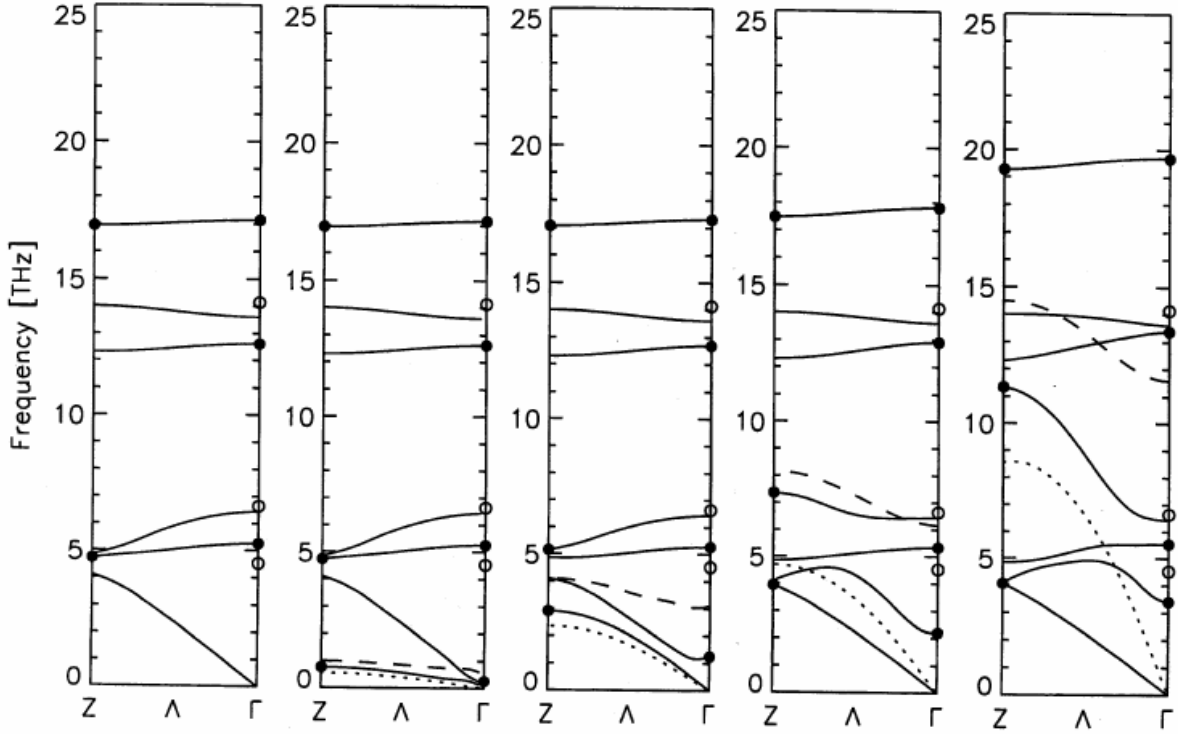

**Fig. 16** Calculated effect of the anisotropy of the structure in terms of the position of the free-plasmon dispersion on the phonon-plasmon scenario along $\Lambda \sim \left(0,0,1\right)$ in *La-Cu-O* as calculated for increasing interlayer couplings represented by the models M1 – M5, from left to right in the figure. The coupling modes at $\Gamma\left(A_{2u}^{LO}\right)$ and at $Z\left(La_z^Z, O_z^Z\right)$ are shown as full dots. The open dots indicate the $A_{2u}^{TO}$ modes at $\Gamma$. $-$: $\Lambda_1$ modes; $---$: free-plasmon branch; $\cdots$: border line for damping due to electron-hole decay.

**Table 4**

Free *c*-axis-plasmon frequency in *THz* at the $\Gamma$ and *Z* point in the BZ according to Eq. (18) for different models with increasing interlayer coupling as explained in the text. The notation CF means modeling with CF's only and CFD represents the same models including, additionally, anisotropic DF's.

|  | $\Gamma$ point | | Z point | |
| --- | --- | --- | --- | --- |
|  | CF | CFD | CF | CFD |
| M2 | 1.26 | 0.62 | 1.27 | 0.99 |
| M3 | 5.05 | 3.11 | 5.26 | 4.09 |
| M4 | 10.16 | 6.16 | 10.44 | 8.16 |
| M5 | 18.75 | 11.58 | 18.30 | 14.49 |

In Fig. 16 the effect of the anisotropy of the material, expressed by the relative position of the free *c*-axis plasmon branch (dashed curve), with respect to the phonon frequencies is shown for different interlayer couplings with increasing magnitude (M1 – M5 from left to right in the figure). In model M1 representing the extreme anisotropic limit of a strictly two-dimensional bandstructure as just mentioned the intraband contribution to the electronic polarizability $\Pi$ vanishes for all wave vectors along the $\Lambda$ direction and the resulting phonon dispersion for the "metallic" phase cannot be distinguished from the dispersion of the insulator. For this reason we call model M1 the "*non-adiabatic insulator*" model. Quantitatively the calculated dispersion curves of M1 are in very good agreement with the inelastic neutron measurement for the insulator $La_2CuO_4$; see [32] and Fig. 5. For very weak interlayer coupling, i. e. a large anisotropy (M2), the plasmon-like $\Lambda_1$ branch is slightly below the free plasmon and consists of two parts of the two lowest $\Lambda_1$ branches. Note, that because of phonon –plasmon mixing, an additional branch appears in the nonadiabatic dispersion of the $\Lambda_1$ modes. So we have seven branches instead of the six symmetry allowed $\Lambda_1$ branches in the adiabatic approximation and for model M1 where the plasmon is soft. For symmetry reasons the three $LO\ A_{2u}^{\Gamma}$ modes and the two $A_{1g}^{Z}$ modes, i. e. the axially symmetric apex-oxygen and *La* breathing modes $O_z^Z$ and $La_z^Z$ (Fig. 14), are allowed to couple to the free plasmon at the $\Gamma$ and the *Z* point, respectively. The latter are shown as full dots in Fig. 16. The open dots indicate the $A_{2u}^{TO}$ modes at $\Gamma$.

Except for the lowest branch (plasmon-like), the optical $\Lambda_1$ branches in M2 (phonon-like modes) are almost indistinguishable from those of the insulator or M1, because in the case of the very low-lying plasmon in model M2 (or M3 too) the slowly oscillating QP cannot screen the phonon-induced bare changes of the long-ranged Coulomb interaction between the ions. The important role of a nonlocal long-ranged EPI in ionic polaron solids has been emphasized in [100]. As a consequence, we will obtain optical activity for the *c*-axis-polarized $A_{2u}$ modes also in the metallic state because the dynamical electronic charge response is much slower than the motions of the ions. Moreover, we get in this highly anisotropic situation large contributions to the nonlocal EPI in terms of the phonon-induced self-consistent changes of the potential $\delta V_\kappa(\vec{q}\sigma,\omega)$, according to Eq. (16), for the phonon-like modes [32]. For example modes like $La_z^Z$ and $O_z^Z$, shown in Fig. 14, where only the atoms in the ionic layers are displaced, generate a strong nonlocal and non-adiabatic coupling to those electrons in the *CuO* planes which are responsible for superconductivity in the cuprates. Such long-ranged nonlocal EPI effects, related to incomplete screening and its dynamic nature, are not possible in conventional isotropic high-density metals and superconductors where a perfect, instantaneous screening of the Coulomb interaction is provided by the large plasmon frequency.

On the other hand, the high-frequency optical lattice vibrations are able to follow the slowly oscillating QP for a low-lying plasmon like in model M2 or M3 leading to some polaronic nature of the latter. Moreover, we obtain a renormalization of the already very low plasmon frequency in form of a coupled plasmon-like mode below the free plasmon; see Fig. 16. We further expect damping effects of the QP via the strong EPI and the induced CF's contributing to decoherence and also via electronic correlation effects not considered in our modelling. Thus, damping or even overdamping of the very low-lying plasmon-like mode becomes very likely. In fact, experimentally the $c$-axis plasmon of metallic $La\text{-}Cu\text{-}O$ seems to be strongly damped or even overdamped for $T > T_c$ [101, 102]. But note, that those experiments are performed at $\vec{q} \simeq \vec{0}$ where the energy of the plasmon is very small, compare with Fig. 17. In recent infrared reflectivity experiments evidence is provided that the c-axis charge transport in LaCuO is intrinsically *coherent* [103] and coherent three-dimensional coupling has been reported in a Tl cuprate superconductor [104].

Increasing the interlayer coupling (models M4 and M5 in Fig. 16), the plasmon-like branch is shifted through the spectrum to higher frequencies. The highest $\Lambda_1$ branch becomes plasmon-like and finally moves out of the spectrum towards the adiabatic limit for even larger interlayer coupling. Simultaneously, the dispersion of the phonon-like $\Lambda_1$ branches develops more and more adiabatic metallic character. In particular we see how the characteristic $\Lambda_1$ branch with the steep dispersion typical for the adiabatic metal (Fig. 4) is formed in model M5. As far as the EPI is concerned we find for the more moderate anisotropy in M4 and M5 large contributions for $\delta V_\kappa$ from the plasmon-like and the phonon-like modes, while for an even less anisotropic regime the plasmon-like mode dominates [32, 46]. Finally, in the adiabatic regime the coupling is strongly reduced.

The scenario of phonon-plasmon mixing can be traced very well at the Z point in Fig. 16 because only the two modes $La_z^Z$ and $O_z^Z$ couple to the free plasmon. For weak interlayer coupling (M2, M3), $La_z^Z$ is mixed with the plasmon while with increasing interlayer coupling the character of the displacements of the plasmon-like mode changes. For model M4 the coupled mode is of mixed $La_z^Z / O_z^Z$ type (at *7.36 THz*) and for M5 or even larger interlayer coupling the $O_z^Z$ mode with the high frequency becomes plasmon-like, finally leaving the range of the phonon spectrum. At the $\Gamma$-point the *LO-TO* splittings of the $A_{2u}^\Gamma$ modes are closed from *below* with an increasing interlayer coupling.

Judging from the experimental evidence, i. e. the insulator-like infrared spectra being practically independent of doping and a strongly damped or even overdamped *c*-axis plasmon, the very anisotropic models M1 or M2 are most appropriate for an approximative description of the situation in *La-Cu-O* in the metallic phase within our theory. In particular, we deduce from our calculations the consistency of the measured (insulator-like) optical *c*-axis data with the calculated (insulator-like) phonon dispersion along the $\Lambda$ direction for the metallic phase in these models. In the case of a stronger, LDA-like interlayer coupling, i. e. a less anisotropic situation, the *LO-TO* splittings get closed – in contrast to the infrared data – because the optical phonons are then screened by the plasmon with a sufficiently high frequency.

A possible explanation of the inconsistency of the current interpretation of the ineleastic neutron scattering data and the optical infrared experiments follows from an investigations of the phonon dispersion in the non-adiabatic region around the $\Lambda$ direction; see Figs. 15 and 17.

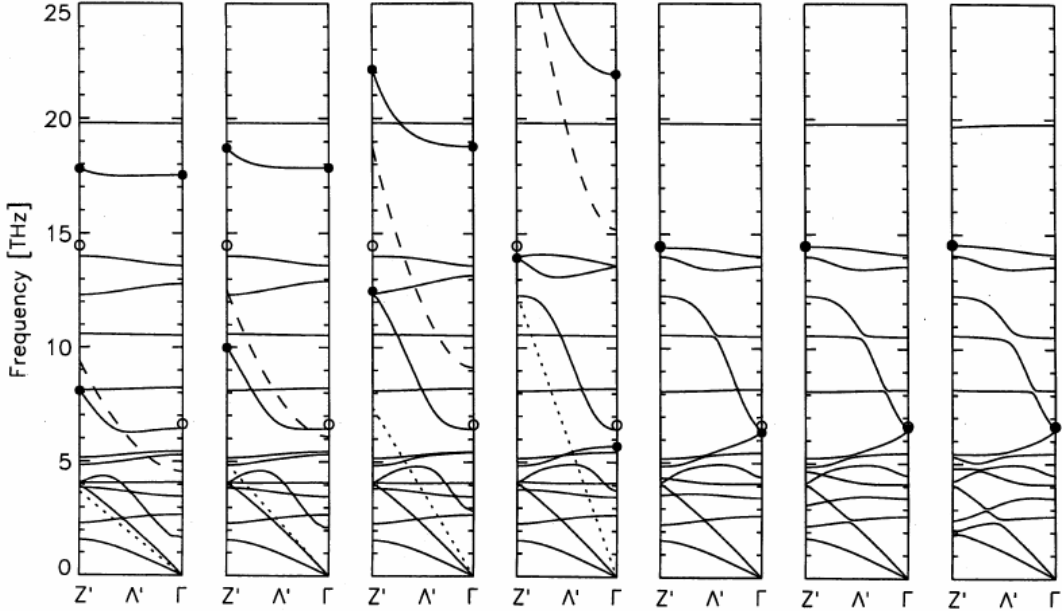

**Fig. 17** Non-adiabatic phonon-plasmon dispersion as calculated with model M1 from $\Gamma$ to $Z' = \left( \varepsilon \frac{2\pi}{a}, 0, \frac{2\pi}{c} \right)$ along the $\Lambda'$ direction (compare with Fig. 15) for different values of $\varepsilon$ and $0 < \zeta \leq 1$ with $\zeta_{min} = 0.01$. From left to right we have in the panels of the figure: $\varepsilon = 0.003, \varepsilon = 0.004, \varepsilon = 0.006, \varepsilon = 0.01, \varepsilon = 0.025, \varepsilon = 0.05, \varepsilon = 0.1$. Only $\Lambda'_1$ branches $(-)$ coupling to the CF's are shown. $---$: free-plasmon branch; $\cdots$: borderline for damping. Dots at $Z' : \bullet O_z^{Z'}(na), o\, O_z^{Z'}(ad)$. Dots at $\Gamma : \bullet A_{2u}^{LO}(ferroelectric; na); o\, A_{2u}^{TO}$ (ferroelectric; ad) na: non-adiabatic; ad: adiabatic.

In neutron scattering experiments a severe complication arises from the finite resolution in momentum transfer perpendicular to the $c$-axis. The wave vector cannot be resolved with sufficient precision along the $\Lambda$ direction. In such a situation a relatively small non-adiabatic insulator-like part of the cone close to the $c$-axis $\left( q_z \gg q_{x,y} \right)$ would be outweighted in the measurement by a significantly larger part where the charge response and the phonon dispersion is nearly adiabatic. Thus, in the neutron experiments, essentially the adiabatic, metallic dispersion with vanishing $A_{2u}$ splittings and the typical $\Lambda_1$ branch with the steep dispersion would show up as seen in Fig. 4.

A representative direction for the cone is the direction $\Lambda' \sim \left( \varepsilon \frac{2\pi}{a}, 0, \frac{2\pi}{c} \right)$ where $\varepsilon$ determines the angle between the $\Lambda$ direction and $\Lambda'$ (Fig. 15) and gives a measure for the size of the cone. Only fourteen phonon branches of $\Lambda_1'$ symmetry couple to the plasmon. In Fig. 17 the non-adiabatic phonon dispersion of these $\Lambda_1'$ branches is calculated with the model M1 for different values of $\varepsilon$ increasing from left to right in the panels of the figure. Increasing $\varepsilon$ rapidly moves the free-plasmon branch (broken curve) out of the phonon spectrum as a consequence of the QP-dispersion in the $\left( k_x k_y \right)$-plane. For $\varepsilon$ between *0.006* and *0.01* we find a region of *crossover* from non-adiabatic insulator-like to adiabatic metallic charge response. The characteristic branch with the steep dispersion appearing in the adiabatic calculations is already seen for $\varepsilon = 0.006$.

The transition from the non-adiabatic to the adiabatic response regime can be traced very well following the $O_z^{Z'}$ mode in Fig. 17. For $\varepsilon \le 0.004$ the lower of the two coupled $O_z^{Z'}$ modes (full dots) is plasmon-like. On increasing $\varepsilon$ successively, this mode rapidly approaches its adiabatic value (open dot), thereby changing its character to phonon-like. Simultaneously, the higher of the two modes, which is for small $\varepsilon$-values phonon- and insulator-like, is shifted out of the spectrum together with the free plasmon and changes its character to plasmon-like. A similar behaviour can be assigned to the longitudinal $A_{2u}$ modes at $\Gamma$. For small $\varepsilon$ we find a typical insulator-like charge response characterized by a large spitting of the "ferroelectric" mode (full and open dots at $\Gamma$). On increasing $\varepsilon$, the *LO-TO* splitting is closed from *below*. The phonon-like $A_{2u}^{LO}$ ("ferroelectric") mode can be identified only for $\varepsilon \approx 0.01$ and larger values, because for smaller ε the lower $A_{2u}^{LO}$ modes are of mixed character. So we have shown in Fig. 17 for $\varepsilon < 0.01$ the higher $A_{2u}^{LO}$ ("ferroelectric") mode only, which becomes plasmon-like for larger $\varepsilon$ and finally leaves the spectrum.

With respect to high-temperature superconductivity it is interesting to remark that the effective interaction between the electrons via phonon-plasmon coupling is attractive in the frequency range between the $A_{2u}^{TO}$ ("*ferroelectric*") mode and the higher $A_{2u}^{LO}$ ("*ferroelectric*") mode. Thus, the large splitting of this mode is favourable for pairing and the role played in this "game" by the imperfectly screened bare long-ranged Coulomb interaction in the non-adiabatic regime around the *c*-axis becomes evident. On the other hand, such a pairing channel is not possible in conventional high-density metals and superconductors. Note in this context that also in recent work [105] it is found that the phonon-plasmon mechanism contributes constructively and significantly to the superconductivity in the cuprates.

We have also investigated in [32] the non-adiabatic phonon-plasmon dispersion along the $\Lambda$" direction (Fig. 15). Such a calculation for a small tube around $\Lambda$ is better adapted to the experimental situation, because the wave vector resolution perpendicular to the $\Lambda$ direction remains constant all the way along the *c*-axis. The discussion of the transition from the non-adiabatic regime for small $\varepsilon$ to the adiabatic regime for larger $\varepsilon$ is similar as for the $\Lambda'$ direction. Again we find a *crossover* from non-adiabatic insulator-like behaviour to adiabatic metallic behaviour for $\varepsilon$ between *0.006* and *0.01*.

From both calculations along $\Lambda'$ and $\Lambda''$ we deduce a narrow region in reciprocal space around the $\Lambda$ direction of non-adiabatic insulator-like charge response. Typically at $\varepsilon \approx 0.006$ metallic behaviour starts to dominate. According to [32] the experimental wave vector resolution perpendicular to $\Lambda$ is on average $\varepsilon \approx 0.03$. As a consequence, the off-axis phonons from the region with a metallic charge response dominate the neutron scattering cross section in the experiments and the measured *c*-axis dispersion looks metallic despite an insulating response behaviour for the phonons propagating strictly along the *c*-axis and within the small non-adiabatic part of the BZ. In this way an explanation of the apparent inconsistency between the current interpretation of the neutron scattering results and the infrared data has been provided, in terms of the phonon-plasmon mechanism.

In the foregoing discussion the size of the non-adiabatic region was based on the extreme anisotropic scenario represented by model M1. An estimate of the reduction of the size of this region with increasing interlayer coupling, i. e. decreasing anisotropy, is shown in Fig. 18, calculating the coupled mode dispersion along the $\Delta'$ direction (Fig. 15) around the *Z* point $\left(0,0,\dfrac{2\pi}{c}\right)$ up to the point $Z' = \left(0.025\dfrac{2\pi}{a},0,\dfrac{2\pi}{c}\right)$ for the models M1, M3 and M5. We extract from Figs. 18 (a – c) that the *acoustic plasmon* gets *massive* for a finite interlayer coupling and that the non-adiabatic region decreases if the material becomes more isotropic. In particular the steep branches in these figure illustrate the non-adiabatic coupling quite well. Taking for example model M1 the steep branch in the middle of Fig 18 (a) changes its character successively from the $La_\zeta^Z$ mode at lower frequency in the non-adiabatic region to the $O_\zeta^Z$ mode at higher frequency in the adiabatic region; compare with Fig. 19, where the corresponding displacement patterns of the modes are shown.

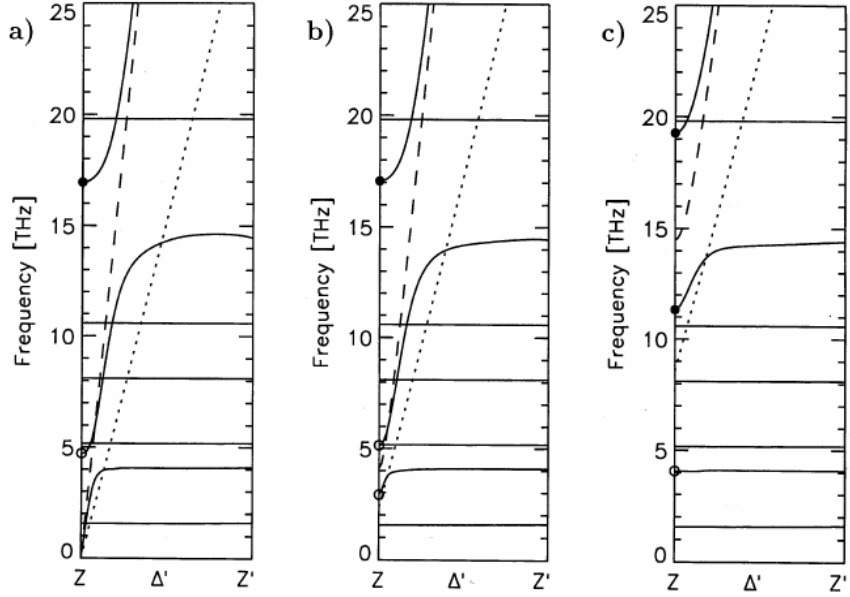

**Fig. 18** Calculated coupled mode dispersion of the $\Delta'_1$ modes $(-)$ from the $Z$ point along $\Delta'$ to

$$Z' = \left(\varepsilon \frac{2\pi}{a}, 0, \frac{2\pi}{c}\right)$$ with $\varepsilon = 0.025$ for the models (a) M1, (b) M3 and (c) M5 from left to right with increasing

interlayer coupling. The coupling modes at $Z$ are: $o\ La_\zeta^Z$ and $\bullet\ O_\zeta^Z$. $- - -$: free-plasmon branch; $\cdots$: borderline for damping.

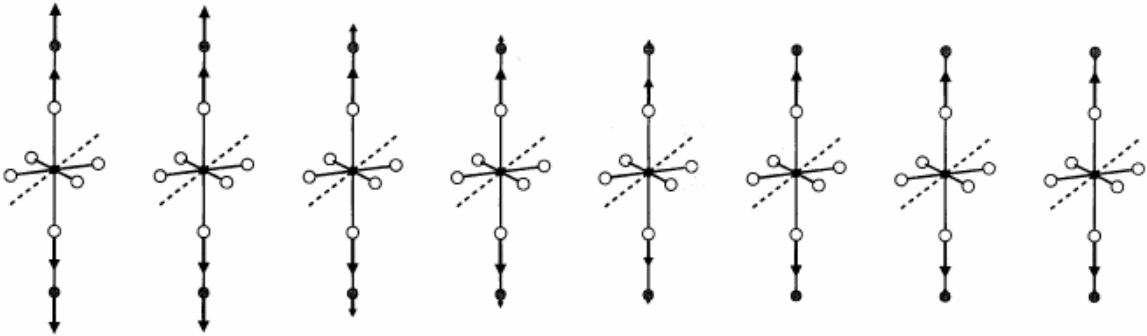

**Fig. 19** Displacement patterns for modes corresponding to the steep branch in the middle of Fig. 18 (a) for $\varepsilon = 0.000, \varepsilon = 0.001, \varepsilon = 0.002, \varepsilon = 0.003, \varepsilon = 0.004, \varepsilon = 0.005, \varepsilon = 0.007, \varepsilon = 0.010$ from left to right,

$$\vec{q}\left(\varepsilon\right) = \left(\varepsilon \frac{2\pi}{a}, 0, \frac{2\pi}{c}\right).$$

Concerning the massive line broadening for the $O_z^Z$ mode, which can also be understood in the phonon-plasmon scenario because of the limited experimental wave vector resolution perpendicular to the $\Lambda$ direction [32], it should be noted that this broadening can be predicted to decrease in proportion to the increasing interlayer coupling with virtually no shift in the frequency. This is because the relevant frequency range sampled by the neutron scattering experiment (steep branch in the middle of Fig. 18) decreases when going from model M1 to M3 while the mode frequency in the dominating adiabatic region is practically not shifted. An enhanced interlayer coupling is realized in the overdoped phase of La-Cu-O [63, 97] where the contribution of the delocalized orbitals to the charge response increases and a dimensional crossover from two to three dimensions is expected to occur in the normal state. As just mentioned this makes the c-axis plasmon more massive and reduces the phase space for strong non-adiabatic coupling. Moreover, with increased delocalization quite generally a weakening of the nonlocal nature of the EPI will take place and a suppression of the antiferromagnet correlations. These features will have important consequences on the nature of the superconductivity that finally disappears because the favouring aspects related to the long-ranged and the short-ranged part of the Coulomb interaction for pairing via combined nonlocal EPI and spin effects are strongly reduced in the overdoped phase.

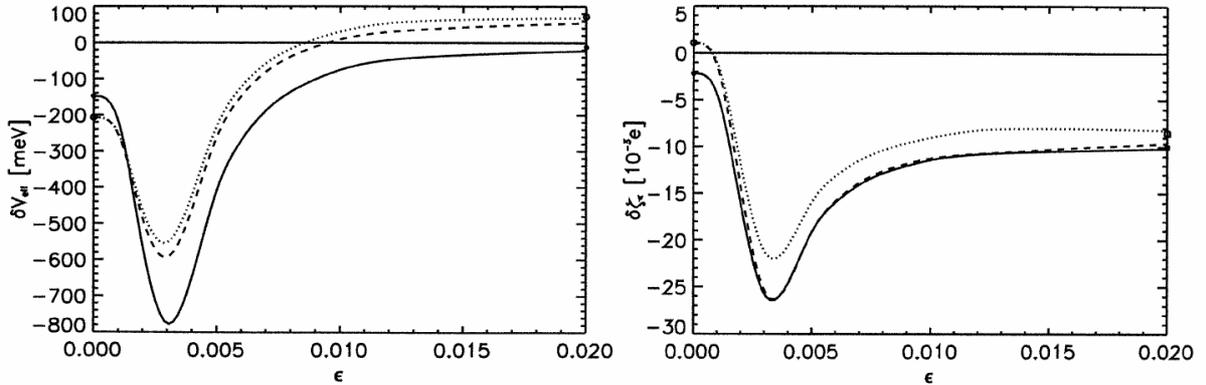

**Fig. 20** Calculated phonon-induced self-consistent change of the non-adiabatic potential $\delta V_{\kappa}\left(\vec{q}\sigma,\omega\right)$ (left) and the corresponding charge-fluctuations $\delta\zeta_{\kappa}\left(\vec{q}\sigma,\omega\right)$ (right) for the steep branch in the middle of Fig. 18 (a) along the $\Delta'$ direction $\left(\varepsilon\frac{2\pi}{a},0,\frac{2\pi}{c}\right), \varepsilon\in\left[0.00,0.02\right]$. *Cu 3d* $(-), O_x-2p\,(\cdots)$ and $O_y-2p\,(---)$ degrees of freedom are considered. The symbols at $\varepsilon=0$ and $\varepsilon=0.02$, respectively, give the corresponding values for the non-adiabatic insulator and the adiabatic metal, respectively.

The transition from the non-adiabatic insulator-like region to the adiabatic metallic region can also be traced by means of the phonon-induced self-consistent changes of the potential an electron feels, $\delta V_\kappa(\vec{q}\sigma,\omega)$ (Eq. (16)), and the corresponding CF's, $\delta\zeta_\kappa(\vec{q}\sigma,\omega)$ (Eq. (12)). These are shown in Fig. 20 for model M1 along the $\Delta'$ direction and the steep branch just mentioned for the *Cu 3d* and the $O_{xy}$ *2p* orbitals [42]. The $\delta V_\kappa$ start at Z with their already large insulator-like values (Fig. 20 left). Then we observe for $\left|\delta V_\kappa\right|$ a strong resonance-like increase in magnitude with a maximum value at about $\varepsilon = 0.003$. This position corresponds to the crossing region in Fig. 18 (a) of the free-plasmon with the $La_z^Z$ mode. By increasing $\varepsilon$ we finally approach the adiabatic coupling regime. Simultaneously the magnitude of $\left|\delta V_\kappa\right|$ strongly decreases and converges to its value for the adiabatic metallic state. The behaviour of the CF's, $\delta\zeta_\kappa$, for the *Cu 3d* and $O_{xy}$ *2p* orbitals is in parallel with that of $\delta V_\kappa$ as can be seen from Fig. 20 (right). The largest CF's in magnitude are again found for the case of resonance.

## 4. Summary and Conclusions

We have developed a reliable microscopic modelling within the framework of linear response theory of the electronic density response, the lattice dynamics, EPI and the dielectric properties of the HTSC's. Our formulation has been applied to the optimally doped metallic, underdoped and insulating phase of the HTSC's in the adiabatic as well as in the non-adiabatic regime. The latter has been shown to be relevant in a certain region around the *c*-axis perpendicular to the *CuO* planes in the metallic state where the screening of the Coulomb interaction is incomplete and the dynamic nature of the EPI becomes essential for this class of highly anisotropic layered materials. Our calculations lead to good structural data for the energy-minimized structure. Complete phonon dispersion curves and particular important phonon modes are obtained for the metallic, underdoped and insulating phase which in general are in quantitative agreement with the experiments. So far different calculations have been performed for $La_2CuO_4$, $Y Ba_2Cu_3O_7$, $Bi_2Sr_2CuO_6$ and $Bi_2Sr_2CaCu_2O_8$.

Starting with an ab initio RIM as a suitable and unbiased reference system for the HTSC's as materials with a strong ionic binding component, non-rigid screening effects and their consequences to phonons and EPI due to localized ionic CF's and DF's have been identified and investigated unambiguously. Ionic CF's controlled by strong Coulomb interactions have been shown to be the dominating electronic polarization process in the *CuO* planes while anisotropic DF's are particularly important for the ions in the ionic layers. Besides the direct

Coulomb interaction, the dielectric coupling via the dipole polarizability of the ions in the layers is an essential ingredient establishing the three-dimensionality of the real materials. In agreement with the experiments we have found in our calculations of the phonon dispersion of *La-Cu-O* characteristic differences during the transition from the insulating to the metallic state upon doping, like the strong softening of the high-frequency OBSM, $\Delta_1/2$ and $O_B^X$, the large renormalization of the symmetrical apex-oxygen breathing mode $O_z^Z$, the unusual behaviour of the axially polarized $\Lambda_1$ branches and $A_{2u}$ modes that can be understood in a phonon-plasmon coupling scenario and other more conventional changes as the closing of the *LO-TO* splittings of the $E_u$ modes polarized parallel to the *CuO* plane. All these effects have been related to and explained by nonlocal EPI effects of CF- and DF type, partly very long-ranged and non-adiabatic. The situation in the underdoped phase of the *p-doped* HTSC's, where a reliable description of the low-energy excitations is missing, has been modelled in terms of a localized charge response at the *Cu* sublattice which is insulator-like and a delocalized, metallic response at the oxygen sublattices of the *CuO* plane, leading to a novel metallic state and a texturing of the electronic structure in the *CuO* layers. Thus, we have insulating regions that coexist with surrounding metallic regions. Reasonable results for the calculated phonon dispersion are obtained. A modelling of the situation for the *n-doped* cuprates also has been proposed expressing the *electron-hole asymmetry*.

Our calculations also have shown correctly the sensitivity of the *HTT* structure with respect to a transition to the *LTO* structure and we have demonstrated that this structural phase transition is driven via the tilt mode essentially by the strong long-ranged ionic forces in the HTSC's. Moreover, we have addressed the dielectric and infrared properties (high-frequency, $\underline{\varepsilon}_\infty$, and static, $\underline{\varepsilon}_0$, dielectric tensor, transverse charges, oscillatorstrengths). Main results are: the matrix elements of $\underline{\varepsilon}_0$ are much larger than those of $\underline{\varepsilon}_\infty$; the infrared response along the *c*-axis and perpendicular to it, each one of them, is dominated in insulating *La-Cu-O* by just one strong phonon reflecting the ionic character of the material along the *c*-axis and the ionic layers, respectively. In particular the *c*-axis response is governed by a "*ferroelectric-like*" mode of $A_{2u}$ symmetry where the *Cu* and *La* cations vibrate coherently against the $O_{xy}$ and $O_z$ anions.

Such strongly coupling infrared active "*ferroelectric-like*" modes also appear in our calculations of the $A_{2u}$ modes in *Bi-2201* and *Bi-2212* cuprate superconductors and are expected to contribute significantly to the low-frequency optical response. We have presented for the *Bi*-compounds detailed calculations of the frequencies and eigenvectors for the infrared active $A_{2u}$ and the $A_{1g}$ Raman modes which may help to clarify the rather controversially discussed assignments in the experimental literature for these modes. From the comparison of the calculated frequencies with the experiments evidence of a metallic behaviour of the *BiO* layers has been provided. This is an important but still open question, because metallic *BiO* planes could play the role of an electron attracting reservoir and therefore may dope the *CuO* layers with holes.

The strong softening of OBSM in the HTSC's upon doping observed by inelastic neutron scattering measurements indicating a strong electron-phonon coupling has been calculated quantitatively and shown to be most likely a generic effect of the *CuO* planes, generated by the nonlocal coupling of the vibrating ions to the localized (ionic) CF's at the *Cu* and $O_{xy}$ and not by Fermi-surface nesting. The size of phase space for the strong nonlocal coupling effects extends to a considerable part of the BZ. An enhancement of softening in the optimally doped metallic phase with largest $T_c$ has been shown to be related to the growing importance of the CF's in the more extended orbitals. In this way a critical degree of delocalization and a corresponding gain in kinetic energy is important for an improvement of superconductivity and the transition temperature in this doping regime and consequently a sufficiently extended set of orbitals including admixtures of higher energy excitation channels seems to be essential to understand the physics of the low energy spectrum in the HTSC's. A possible behaviour to be expected for the modes in the overdoped phase has also been addressed. There are higher energy CF's related to the localized *Cu 3d* states and lower energy CF's related to the more delocalized *O 2p* and *Cu 4s (4p)* states and a critical mixing of both is required to achieve the high $T_c$ values in the optimally doped material. Increasing doping the contribution of the low-energy CF's of the delocalized states is enhanced, the EPI becomes more local in character and the many-body effects leading to antiferromagnetic SF's are expected to weaken. The latter ultimately die out and a more conventional Fermi-liquid state will be restored at high doping. Thus, the degree of itinerancy of the electrons, mainly via *Cu 4s*, critically controls the degree of nonlocallity of the EPI, the crossover between the different regimes of the electronic structure and the experimentally observed $T_c$. In *Y-Ba-Cu-O* the study of the anomalous dispersion of the OBSM is complicated by anticrossing effects with other modes of the same symmetry type which makes a speculative interpretation of the experimentally observed branch splitting as a signature of charge stripes unlikely. We also have explained the anisotropic behaviour of softening along the (1,0,0) and (1,1,0) directions, respectively, in terms of CF's excited additionally at the oxygen ions in case of the $\Delta_1/2$ modes not possible for the $O_B^X$ mode. Moreover, we have illustrated by means of the OBSM how coupled lattice-, charge (orbital)- and spin-degrees of freedom may act synergetically to enhance the CF's and the EPI as well as the SF's. So, all relevant degrees of freedom of the system can work together constructively in the optimally doped regime in the pairing mechanism of the cuprates. The resulting phonon-induced charge redistribution for the $\Delta_1/2$ and the $O_B^X$ mode appears as a dynamic charge ordering in the form of localized stripes of alternating sign in the *CuO* plane.

We also have addressed what we think to be essential factors which modify the charge response and maximize $T_c$ in the multilayer high-temperature materials as compared to the single layer case .

For the *K*-doped *non-cuprate perovskite* high-temperature superconductor *Ba-Bi-O* our calculations predict a very strong and anisotropic phonon softening for the OBSM essentially induced by nonlocal coupling of the oxygen displacements to the *Bi 6s* CF's.

Finally, we have investigated interlayer phonons along the *c*-axis and their accompanying charge response. Our calculations for *La-Cu-O* strongly support a non-adiabatic coupled *phonon-plasmon scenario* in a certain region around the *c*-axis which is due to the dynamic nature of the screened interaction. Such a scenario, in contrast to a response description within the adiabatic approximation, is consistent with the observed strong *c*-axis infrared activity in the doped metallic state, and allows for an explanation of the apparent inconsistency between the current interpretation of the neutron scattering results and the infrared data. It explains the large softening and the massive line broadening of the $O_z^Z$ mode seen in the experiments during the transition from the insulating to the metallic phase. Changes of the mode behaviour have been predicted for the overdoped phase. The size of the non-adiabatic region around the *c*-axis where the EPI is controlled by dynamical screening has been investigated in some detail. It depends on both, the strength of interlayer coupling and the electronic structure in the *CuO* plane pointing to an interplay of non-adiabatic effects and strong electronic correlations. So, the charge dynamics nearby the *c*-axis is at least partially related to the electronic state of the *CuO* planes and vice versa.

Because of the significantly reduced screening perpendicular to the *CuO* planes strong nonlocal, non-adiabatic long-ranged EPI effects have been found in this region which are favourable for pairing in the HTSC's in both, the phonon-like and the plasmon-like channel. This is not at all possible in conventional high-density metals and superconductors because of a perfect instantaneous screening of the Coulomb interaction. A particularly large resonance-like enhancement due to the long-ranged Coulomb contribution of the interaction to the EPI is found in the crossing region of the uncoupled (free) plasmon branch with the corresponding uncoupled phonon branch providing an extra channel for the formation of Cooper pairs. Altogether from our findings of a strong nonlocal EPI via CF's in particular for the generic OBSM which can be dealt within adiabatic approximation and the non-adiabatic coupled phonon-plasmon modes it becomes evident that these modes and the CF's made visible in our calculations, i. e. incoherent CF's in context with the OBSM and coherent CF's in case of the *c*-axis modes, give an important contribution to the pairing mechanism in the HTSC's based on lattice- and charge-degrees of freedom, synergetically enhanced by corresponding SF's in the *CuO* plane. It also has been outlined, that the OBSM may play a role for the antiferromagnetic spin resonant excitation that develops in the superconducting state. From recent experiments, however, it can be concluded that high-transition temperature superconductivity seems to be present in the absence of the magnetic resonance mode. Interestingly, the OBSM are largely controlled by the short-ranged part of the Coulomb interaction and the phonon-plasmon modes by the long-ranged part. In this way both parts of the Coulomb interaction participate in a characteristic way in pairing and determine its symmetry. Anisotropic pairing is possible because of the nonlocality of the interaction in space, like the predominantly *d-wave symmetry* found in many experiments in the underdoped and optimally hole-doped HTSC's. However, it seems not unlikely that in parallel with the increased delocalization of the electronic states in the *CuO* plane upon doping as suggested by our investigations, in particular at *Cu*, a *s-symmetry* component may start to develop in the overdoped materials.

Further evidence for the relevance of the EPI by means of the OBSM and the plasmons for the superconductivity in the cuprates has been provided by other work in the field. So it has been argued that the $\Delta_1 / 2$ mode strongly contributes to d-wave pairing and that low-energy plasmons contribute constructively and in a significant way to the superconductivity. Quite generally, our results of a strong nonlocal EPI leading to a considerable mixing of electron- and phonon-degrees of freedom in form of significant dynamic charge modulations indicate that the nature of many-body effects and superconductivity in the cuprates is associated also with specific CF's and phonons and not only with electron-electron interactions that lead to modulations in the spin channel as has been commonly assumed. Obviously, from the high-temperature superconductivity found in the experiments and the discussion in this paper superconductivity is an attractive possibility of a correlated electron system with strong nonlocal EPI favoured by the ionic component of binding.

**Acknowledgements**

I would like to thank G.A. Hoffmann, M. Klenner, F. Schnetgöke and Q. Chen for the fruitful collaboration during the last years and T. Kuhn for a careful reading of the manuscript.